\documentclass[12pt]{article}
\usepackage{amsmath,amssymb}
\usepackage{bm}
\usepackage{color}
\usepackage{graphicx}
\usepackage{cite}
\usepackage{mathtools}
\usepackage{breqn}

\usepackage{here}
\usepackage{comment}

\usepackage{amsthm}
\newtheoremstyle{break}
{\topsep}{\topsep}%
{\normalfont}{0pt}%
{\bfseries}{}%
{\newline}{}
\theoremstyle{break}

\newtheorem*{thmhom}{Fundamental homomorphism theorem}
\newtheorem*{thmcor}{Correspondence theorem}
\newtheorem{thmiso}{Isomorphism theorem}
\newtheorem*{thmAbel}{Fundamental structure theorem of finite Abelian group}
\newtheorem*{thmcirc}{The Chinese remainder theorem}
\newtheorem*{pf}{Proof}

\newtheorem{step}{Step}

\begin{document}

\title{
\begin{flushright}
\ \\*[-80pt]
\begin{minipage}{0.2\linewidth}
\normalsize
%arXiv:YYMM.NNNN \\
EPHOU-21-019\\*[50pt]
\end{minipage}
\end{flushright}
% Title
{\Large \bf
Anomaly of non-Abelian discrete symmetries 
\\*[20pt]}}
% /Title

\author{
%Shota Kikuchi$^{a}$,
%\footnote{B's mail}
Tatsuo Kobayashi and
%\footnote{C's mail}
%Kaito Nasu$^{a}$, 
%\footnote{D's mail}
Hikaru Uchida
%\footnote{E's mail}
\\*[20pt]
\centerline{
\begin{minipage}{\linewidth}
\begin{center}
{\it \normalsize
Department of Physics, Hokkaido University, Sapporo 060-0810, Japan} \\*[5pt]
\end{center}
\end{minipage}}
\\*[50pt]}

\date{
\centerline{\small \bf Abstract}
\begin{minipage}{0.9\linewidth}
\medskip
\medskip
\small
We study anomalies of non-Abelian discrete symmetries; 
which part of non-Abelian group is anomaly free and which part 
can be anomalous.
It is found that the anomaly-free elements of the group $G$ 
generate a normal subgroup $G_0$ of $G$ and the residue class group $G/G_0$, which becomes the anomalous part of $G$, is isomorphic to a single cyclic group.
The derived subgroup $D(G)$ of $G$ is useful to study the anomaly structure.
This structure also constrains the structure of the anomaly-free subgroup; the derived subgroup $D(G)$ should be included in the anomaly-free subgroup.
We study the detail structure of the anomaly-free subgroup from the structure of the derived subgroup in various discrete groups.
For example, when $G=S_n \simeq A_n \rtimes Z_2$ and $G=\Delta(6n^2) \simeq \Delta(3n^2) \rtimes Z_2$, in particular, $A_n$ and $\Delta(3n^2)$ are at least included in the anomaly-free subgroup, respectively.
This result holds in any arbitrary representations.
\end{minipage}
}

\begin{titlepage}
\maketitle
\thispagestyle{empty}
\end{titlepage}

\newpage

% ------------------------------------------------------ %
% ------------------------------------------------------ %
% ------------------------------------------------------ %
% ------------------------------------------------------ %

\section{Introduction}
\label{sec:Intro}

Symmetries are significant to understand various physical phenomena.
The standard model of particle physics can explain the strong and electroweak interactions by the gauge $SU(3) \times SU(2) \times U(1)$ symmetry.
Other continuous global and local symmetries are also important in particle physics, e.g., 
baryon and lepton number symmetries.
In addition, Abelian and non-Abelian continuous symmetries are often used for model building.

Not only continuous symmetries, but also discrete symmetries are important.
C, P, and T are essential discrete symmetries in particle physics.
The R-parity is useful to forbid the fast proton decay in minimal supersymmetric 
standard model.
The $Z_N$ symmetry can be used to stabilize dark matter candidates.
The origin of the quark and lepton flavor structures is one of the most significant mysteries 
in particle physics.
Indeed, a lot of studies have been carried out.
Among them one of most interesting approaches is to impose  non-Abelian discrete flavor symmetries~\cite{
	Altarelli:2010gt,Ishimori:2010au,Ishimori:2012zz,Hernandez:2012ra,
	King:2013eh,King:2014nza}
on the generations of quarks and leptons.
%Thus, not only continuous symmetries but also discrete symmetries are keys to understand backgrounds of phenomena.

However, even if there exists a symmetry in a theory at classical level, the symmetry can be broken by quantum effects, that is, the anomaly.
In the case of continuous symmetry, global chiral $U(1)$ symmetry, for example, can be broken by the chiral anomaly unless the sum of $U(1)$ charges of chiral fermions vanishes.
Fujikawa's method~\cite{Fujikawa:1979ay,Fujikawa:1980eg} is a useful way to compute such chiral anomalies .
%Such chiral anomalies can be calculated in the Fujikawa's method~\cite{Fujikawa:1979ay,Fujikawa:1980eg}.

Similarly, anomalies of the Abelian discrete symmetry $Z_N$ were studied in 
\cite{Krauss:1988zc,Ibanez:1991hv,Banks:1991xj}.
Furthermore, anomalies of non-Abelian symmetries were studied 
 by using the Fujikawa's method in Ref.~\cite{Araki:2008ek}.
Each element $g$ of the non-Abelian group $G$ generates the Abelian discrete symmetry 
$Z_N$ when $g^N= e$, where $e$ is the identity.
If all of these Abelian symmetries corresponding to all the elements $g$ in $G$ 
are anomaly free, then the full symmetry $G$ is anomaly free.
When some parts are anomalous, unbroken symmetry corresponds to the 
subgroup of $G$, which does not include anomalous elements.
%In particular,
Indeed, Refs.~\cite{Ishimori:2010au,Ishimori:2012zz} show which elements are anomaly free 
or can be anomalous for various examples of discrete groups.
In addition, Ref.~\cite{Chen:2015aba} shows the anomaly-free condition by introducing the derived subgroup of $G$, $D(G)$.
Here, the derived subgroup is generated by commutator elements, $xyx^{-1}y^{-1}$, where $x$ and $y$ are elements of $G$.
In particular, it shows that perfect groups, which are defined as $D(G)=G$, are perfectly anomaly free.
In this paper, we study more detail anomaly structure of non-Abelian discrete symmetries.
%In this paper, we study more about anomalies of non-Abelian discrete symmetries.
%Note that they are discussed in each group and each representation.
%In addition, the explicit anomaly free subgroups have not been studied.
%That is, our purpose of this paper is to understand the anomaly structure of 
%non-Abelian discrete symmetries deeply. 
%That is, our purpose of this paper is to understand the detail structure of anomaly-free subgroups and anomalous parts of non-Abelian discrete symmetries.
%We explore generic structure of anomaly-free and anomalous parts 
%of non-Abelain discrete groups.
That is, we explore the detail structure of anomaly-free subgroups and anomalous parts of non-Abelian discrete symmetries generically.

This paper is organized as follows.
In section~\ref{sec:Anomalyrev}, we review anomalies of (discrete) symmetry.
In particular, a group element whose determinant is trivial becomes an anomaly-free transformation.
In section~\ref{sec:rep}, we study anomaly-free and anomalous structure of a discrete group, explored from the determinant of a representation.
We have found that the anomaly-free elements construct a normal subgroup of the discrete group and the residue class group, which becomes the anomalous part, is isomorphic to a single cyclic group.
This structure is important to explore the structure of the anomaly-free subgroup even if we do not specify the representation.
The derived subgroup is the smallest normal subgroup, which leads the residue class group isomorphic to an Abelian group including a single cyclic group. 
This structure constrains the structure of the anomaly-free subgroup; the derived subgroup $D(G)$ should be included in the anomaly-free subgroup.
Then, we study the detail structure of the anomaly-free subgroup through various examples of discrete groups in section~\ref{sec:group}.
In section~\ref{sec:comment}, we comment on generic theories.
In section~\ref{sec:conclusion}, we conclude this study.
In Appendix~\ref{app:isomorphism}, the fundamental homomorphism and then the isomorphism theorems as well as the correspondence theorem are arranged.
In Appendix~\ref{app:semidirect}, some properties related to semidirect products are given.
In Appendix~\ref{app:Abelian}, some properties of finite Abelian groups are given.

% ------------------------------------------------------ %
% ------------------------------------------------------ %
% ------------------------------------------------------ %
% ------------------------------------------------------ %

\section{Anomalies of discrete symmetry}
\label{sec:Anomalyrev}

First, we briefly review anomalies of discrete symmetry~\cite{Araki:2008ek}.
Let us assume that a classical action $S$ with a set of chiral fermions $\psi_L = P_L \Psi$ is invariant under unitary  transformation for the fermions, $\psi_L \rightarrow \rho(g) \psi_L,\ \forall g \in G$, where $G$ is a group and $\rho(g)$ is a unitary representation of $g \in G$.
In this case, we say that the theory, at least at classical level, has chiral $G$ symmetry.
For example, in the case of $G=Z_N$ symmetry, the generator $g \in Z_N$ satisfies $g^N=e$ and the unitary representation can be expressed as $\rho(g)_{jk}=e^{i\alpha q_j}\delta_{jk}$ with the phase parameter, $\alpha=2\pi/N$, and the $Z_N$ charge of $j$ th component of $\psi_L$, $q_j \in \mathbb{Z}/N\mathbb{Z}$.

However, such a classical chiral symmetry can be broken at quantum level.
First, let us see the global $G=Z_N$ symmetry under background non-Abelian gauge fields as well as gravity.
We denote the gauge group by $G_{\rm gauge}$ and the fermions have a representation ${\bf R}$ 
under $G_{\rm gauge}$.
In the Fujikawa's method~\cite{Fujikawa:1979ay,Fujikawa:1980eg}, in particular, the measure in the path integral, $\int D\Psi D\bar{\Psi} e^{iS}$, can transform as
\begin{align}
D\Psi D\bar{\Psi} \rightarrow J(\alpha) D\Psi D\bar{\Psi},
\end{align}
%under $g (\in G)$ transformation, where $J(g)$ denotes the Jacobian of $g$ transformation.
%Here, we focus on the case that $G$ is a global symmetry.
%We assume that the fermions interact with non-Abelian gauge fields as well as 
%gravity.
%We denote the gauge group by $G_{\rm gauge}$ and the fermions have a representation ${\bf R}$ 
%under $G_{\rm gauge}$.
%First, we consider the case that $G$ is Abelian, $Z_N$, and its generator is $g$, i.e. $g^N=e$.
%Fermions have $Z_N$ charges, 
%$q_i$.
%Then,
where the Jacobian can be written as~\cite{Alvarez-Gaume:1983ihn,Alvarez-Gaume:1984zlq},
\begin{align}
J(\alpha) = {\rm exp}\left[ i\int d^4x ~(A(x;\alpha)_{\rm gauge} + A(x;\alpha)_{\rm grav})\right],
\end{align}
with $\alpha=2\pi/N$.
The anomaly functions are written by 
\begin{align}
A(x;\alpha)_{\rm gauge} =\frac{1}{32\pi^2} \epsilon^{\mu\nu\rho\sigma} {\rm Tr}(\alpha q_j [F_{\mu\nu} F_{\rho\sigma}]),  
\end{align}
where ${\rm Tr}$ denotes the summation over all internal indices, 
and 
\begin{align}
A(x;\alpha)_{\rm grav}=-\frac{1}{2} \frac{1}{384\pi^2} \frac{1}{2} \epsilon^{\mu\nu\rho\sigma} R_{\mu\nu}^{\lambda\gamma} R_{\rho\sigma\lambda\gamma} {\rm tr} (\alpha q_j{\bf R}).
\end{align}
The index theorems~\cite{Alvarez-Gaume:1983ihn,Alvarez-Gaume:1984zlq} imply
% and the fermions feel background gravity and gauge field of a group $G'$ which commutes with the group $G$.
%In this background, by considering the index theorems~\cite{Alvarez-Gaume:1983ihn,Alvarez-Gaume:1984zlq}, which imply
\begin{align}
\int d^4x \frac{1}{32\pi^2} \epsilon^{\mu\nu\rho\sigma} F_{\mu\nu}^a F_{\rho\sigma}^b
{\rm tr}[t^a t^b] \in \mathbb{Z}, \quad
\frac{1}{2} \int d^4x \frac{1}{384\pi^2} \frac{1}{2} \epsilon^{\mu\nu\rho\sigma} R_{\mu\nu}^{\lambda\gamma} R_{\rho\sigma\lambda\gamma} \in \mathbb{Z},
\label{eq:index}
\end{align}
where $t^{a,b}$ denote generators of $G_{\rm gauge}$ in the ${\bf R}$ representation.
We use the normalization of Dynkin index $T_2({\bf R})$, 
\begin{align}
T_2({\bf R}) \delta_{ab}= {\rm tr}[t^a t^b] ,
\end{align}
such that $T_2({\bf R})=1/2$ for $N$ fundamental representation of  $SU(N)$ 
and $T_2({\bf R})=1$ for $2N$ vector representation of  $SO(2N)$.
For example, we have $T_2({\bf R})=3$ for {\bf 27} representation of $E_6$.
Thus, the anomaly-free condition for the mixed anomaly $Z_N-G_{\rm gauge}-G_{\rm gauge}$ 
is obtained as 
\begin{align}
J(\alpha) = e^{2\pi i \sum q_j 2T_2({\bf R})n/N} = 1, \ \forall n \in \mathbb{Z} \quad \Leftrightarrow \quad
\sum q_j 2T_2({\bf R}) \equiv 0\ ({\rm mod}\ N).
\end{align}
Otherwise, the $Z_N$ symmetry can be anomalous.

Next, let us see the global non-Abelian discrete symmetry $G$.
%For each element $g$ of the group $G$, 
We can study its anomalies similarly as the $Z_N$ symmetry case
since each element $g$ of the group $G$ satisfies $g^{N(g)}=e$,
where $N(g)$ is the order of $g$.
In general, 
fermions construct a multiplet under the non-Abelian symmetry $G$.
For such a multiplet, the unitary representation of $g \in G$, $\rho(g)$, forms as unitary matrix. 
%each element $g$ of the group $G$ is represented by 
%a unitary matrix, $\rho(g)$.
Here, we can always make the $\rho(g)$ diagonalized as $\rho(g)_{jk}=e^{i\alpha(g)q_j(g)}\delta_{jk}$
with the phase parameter of the $g$ transformation, $\alpha(g)=2\pi/N(g)$, and the charge of $j$ th component of the multiplet for $g$ transformation, $q_j \in \mathbb{Z}/N(g)\mathbb{Z}$, by taking the appropriate base of the fermions.
Note that, in such a base, the unitary matrices of some of the other elements $g' \in G$, $\rho(g')$, become non-diagonalized matrices.
Then, we can apply the analysis of the anomalies of the $Z_N$ symmetry to the anomalies of the non-Abelian discrete symmetry $G$.
%Then, the determinant ${\rm det}\rho(g)$ is important to 
%evaluate anomalies.
Suppose that fermion multiplets correspond to the representation 
$\rho(g)$ of the non-Abelian discrete symmetry $G$ 
and representations ${\bf R}$ of the non-Abelian gauge symmetry $G_{\rm gauge}$.
Then, the anomaly-free condition for the mixed anomalies 
$G-G_{\rm gauge}-G_{\rm gauge}$ is written by 
\begin{align}
J(\alpha(g)) = e^{2\pi i \sum q_j 2T_2({\bf R})n/N(g)} = (e^{2\pi i Q(g)/N(g)})^{\sum_{\bf R}2T_2({\bf R})n} = ({\rm det}\rho(g))^{\sum_{\bf R}2T_2({\bf R})n} = 1, \ \forall n \in \mathbb{Z},
\end{align}
where $Q(g) \equiv \sum_{j} q_j(g)$ and this $Q(g)$ is preserved even if the representation $\rho(g)$ is not diagonalized.
Hence, we can say the symmetries including only the elements $g$ corresponding to 
\begin{align}
{\rm det}\rho(g)=1, \quad (\Leftrightarrow  Q(g) \equiv 0\ ({\rm mod}\ N(g))),
\end{align}
are always anomaly free.
Other parts in $G$ can be anomalous.
The anomalies of symmetries corresponding to elements $g$ with 
$\rho(g) \neq 1$ depend on matter contents.
That is, for $\sum_{\bf R}2T_2({\bf R})=M$, 
the subgroup constructed by elements $g$ with 
$({\rm det}\rho (g))^M =1$ is anomaly free, although 
the subgroup constructed by elements $g$ with 
${\rm det}\rho (g) =1$ is always anomaly free.
Thus, the determinant ${\rm det}\rho (g)$ is the key point 
in the analysis of following sections.

%the Jacobian $J(g)$ can be expressed as
%\begin{align}
%J(g) = e^{in\alpha(g)Q(g)} = ({\rm det}\rho(g))^n,
%\end{align}
%where $Q(g) \equiv \sum_{j}q_j(g)$ and $n \in \mathbb{Z}$ comes from Eq.~(\ref{eq:index}).
%Note that for the $G-G'-G'$ anomaly, the components of fermions in a same multiplet of $G'$ are summed only once.
%Thus, when ${\rm det}\rho(g)=1$, $g$ transformation is invariant even at quantum level, that is called an anomaly free transformation.
%Otherwise, $g$ can be broken at quantum level, that is called chiral anomaly.
%Here, in the case of a continuous $G$ symmetry, ${\rm det}\rho(g)=1$ is satisfied only if $Q(g)=0$.
%On the other hand, in the case of a discrete $G$ symmetry, since $g^{N(g)}=e$, $\alpha(g)=2\pi/N(g)$, $Q(g) \in \mathbb{Z}$, and then ${\rm det}\rho(g)=1$ is satisfied if $Q(g) \equiv 0\ ({\rm mod}\ N(g))$, where $N(g)$ denotes the order of $g$.
%In the following analysis, we study anomaly free subgroups of discrete groups $G$ in terms of ${\rm det}\rho(g)\ (g \in G)$.

% ------------------------------------------------------ %
% ------------------------------------------------------ %
% ------------------------------------------------------ %
% ------------------------------------------------------ %

\section{Anomaly structure explored from determinant of representations}
\label{sec:rep}

In this section, we study the anomaly structure by use of concrete 
representations $\rho(g)$ for elements $g$ in the non-Abelian symmetry $G$.
%Suppose that $g^{N(g)}=e$.
Here, we concentrate mainly on the theory with $\sum_{\bf R}2T_2({\bf R})=1$ and the anomaly-free condition 
 ${\rm det}\rho (g) =1$.
However, it is straightforward to extend our analysis to  the theory with $\sum_{\bf R}2T_2({\bf R})=M > 1$ and 
anomaly-free condition $({\rm det}\rho (g))^M =1$.

\begin{figure}[H]
\centering
\includegraphics[bb=0 0 650 510,width=9cm]{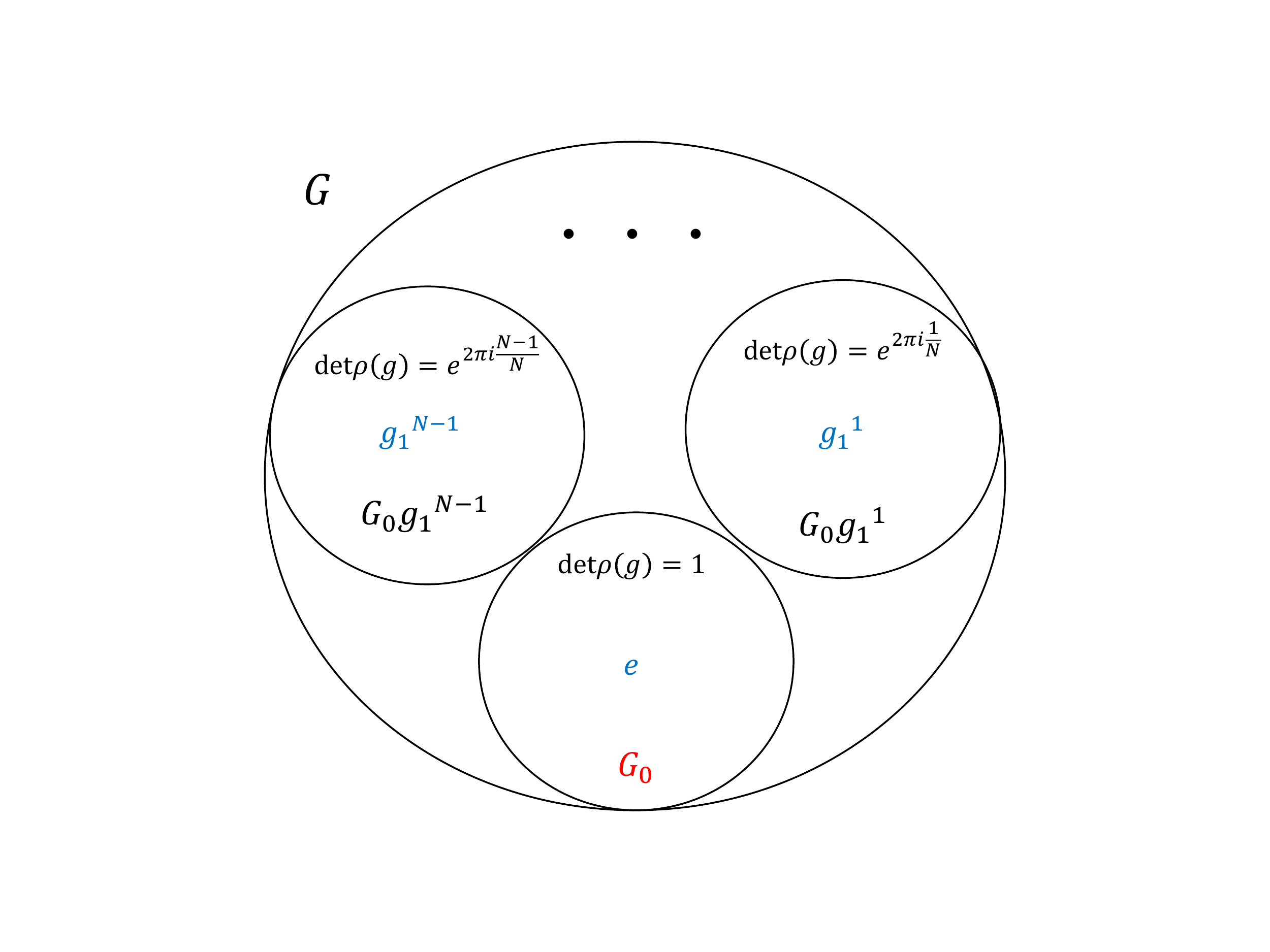}
\caption{Image of cosets $G_0g_1^k$ whose elements $g$ satisfy ${\rm det}\rho(g)=e^{2\pi ik/N}$. The representative element of each coset is written by blue.}
\label{fig:anomalousZN}
\end{figure}

%In this section, we consider the case that we know the representation of $\forall g \in G$, $\rho(g)$.
Given representations of $\forall g \in G$, $\rho(g)$, 
we can calculate ${\rm det}\rho(g)$ explicitly.
Suppose that $g^{N(g)}=e$ and then ${\rm det}\rho(g)^{N(g)}=1$ for a fixed element $g$, and also
 $({\rm det}\rho(g))^{N}=1$ for  any element $g$ in $G$. 
Then, we can write all of them as ${\rm det}\rho(g)=e^{2\pi i Q'(g)/N}\ (\forall g \in G)$, where $Q'(g)$ is given by $Q'(g)=Q(g)N/N(g)$.
As shown in the previous section, if ${\rm det}\rho(g)=1\ (Q'(g) \equiv 0\ ({\rm mod}\ N))$, 
the element $g$ corresponds to  anomaly-free  transformation.
Then, we define
\begin{align}
G_0 \equiv \{ g_0 \in G | {\rm det}\rho(g_0)=1 \} ,
\end{align}
as the subset of $G$.
From the following proof, we can find $G_0$ becomes a normal subgroup of $G$, $G_0 \triangleleft G$.
Thus, if anomalous transformations are fully broken by quantum effects, the symmetry $G$ is broken to the normal subgroup $G_0$ at quantum level.
\begin{pf}
We can prove that $G_0$ is a subgroup of $G$, $G_0 \subset G$, from (I) and then $G_0$ is also a normal subgroup of $G$, $G_0 \triangleleft G$, from (II).
\begin{enumerate}
\renewcommand{\labelenumi}{(\Roman{enumi})}
\item When we take $\forall g_0 \in G_0$ and $\forall g'_0 \in G_0$ (${\rm det}\rho(g_0)={\rm det}\rho(g'_0)=1$), 
the element $g_0g'_0$ is also included in $G_0$, $g_0g'_0 \in G_0$ (${\rm det}\rho(g_0g'_0)=1$). In particular, the identity element $e$ is included in $G_0$ (${\rm det}\rho(e)=1$), and also when we take $\forall g_0 \in G_0$ (${\rm det}\rho(g_0)=1$), the inverse element $g_0^{-1}$ is included in $G_0$ (${\rm det}\rho(g_0^{-1})=1$).
\item When we take $\forall g_0 \in G_0$ (${\rm det}\rho(g_0)=1$) and $\forall g \in G$, 
the conjugate element $gg_0g^{-1}$ is also included in $G_0$, $gg_0g^{-1} \in G_0$ (${\rm det}\rho(gg_0g^{-1})={\rm det}\rho(g_0)=1$).
\end{enumerate}
\end{pf}
Now, we can rewrite $\forall g \in G$, which satisfies ${\rm det}\rho(g) = e^{2\pi i k/N}\ (Q'(g) \equiv k\ ({\rm mod}\ N))$ as $g=g_0g_1^k$, where $g_1$ satisfies ${\rm det}\rho(g_1) = e^{2\pi i/N}\ (Q'(g_1) \equiv 1\ ({\rm mod}\ N))$ and $\exists g_0 \in G_0$.
In other words, the coset, whose element $g$ satisfies  ${\rm det}\rho(g) = e^{2\pi i k/N}$, can be expressed as $G_0g_1^k$. (See also Fig.~\ref{fig:anomalousZN}.)
Actually, from the following proof, we can find such cosets consist the residue class group $G/G_0 \simeq Z_N$.
%\footnote{If we define $G_n \equiv \{ g \in G | {\rm det}\rho(g) = e^{2\pi i Q'(g)/N} = e^{2\pi i Q''(g)/n},\ Q'(g)(=Q(g)N/N(g))=Q''(g)N/n \}$, $G_n$ is also a normal subgroup of $G$, $G_n \triangleleft G$, and it includes of $G_0$, $G_0 \subset G_n$, which also means $G_0 \triangleleft G_n$. In this case, we can similarly find $G/G_n \simeq Z_{N/n}$ and $G_n/G_0 \simeq Z_n$. Indeed, by use of the isomorphism theorem $3$ in Appendix~\ref{app:isomorphism}, we find that $G/G_n \simeq (G/G_0)/(G_n/G_0) \simeq Z_N/Z_n \simeq Z_{N/n}$.}.
\begin{pf}
We can prove that the residue class group $G/G_0$ is Abelian from (I) and also isomorphic to $Z_N$ from (II).
\begin{enumerate}
\renewcommand{\labelenumi}{(\Roman{enumi})}
\item
$G_0g_1^{k_1}$ and $G_0g_1^{k_2}$ satisfy the relation $(G_0g_1^{k_1})(G_0g_1^{k_2}) = (G_0g_1^{k_2})(G_0g_1^{k_1}) = G_0g_1^{k_1+k_2}$.
\item
$G_0g_1^{N-k}$ becomes inverse coset of $G_0g_1^{k}$ since $g_1^N \in G_0$ (${\rm det}\rho(g_1^N)=1$).
\end{enumerate}
\end{pf}
Here, the element $g_1 \in G$ generally satisfies $g_1^N = g_0,\ \exists g_0 \in G$ while it satisfies $g_1^{N(g_1)}=e$.
Then, we find $G_0 \cap Z_{N(g_1)} = Z_{N(g_1)/N}$ in general, where $Z_{N(g_1)}$ is the subgroup of $G$ generated by $g_1$ and $Z_{N(g_1)/N}$ is the subgroup of $G_0$ generated by $g_1^N$.
Actually, by using the isomorphism theorem $2$ in Appendix~\ref{app:isomorphism}, we can also obtain
\begin{align}
G/G_0 \simeq G_0Z_{N(g_1)}/G_0 \simeq Z_{N(g_1)}/Z_{N(g_1)/N} \simeq Z_N. \notag
\end{align}

If $g_1$ satisfies $g_1^N=e \in G_0\ (N(g_1)=N)$, in particular, $g_1$ generates $Z_N$ subgroup of $G$ and the $Z_N$ subgroup satisfies $G=G_0Z_N$ and $G_0 \cap Z_N = \{e\}$.
Thus, in this case, $G$ can be decomposed\footnote{See also Appendix~\ref{app:semidirect}.} as
\begin{align}
G \simeq G_0 \rtimes Z_N. \label{eq:G0ZN}
\end{align}
It means that the anomaly-free and anomalous parts of $G$ can be separated.
%the following subset of $G$,
%\begin{align}
%H_N = \{ e,\ g_1^k\ (k=1,2,...,N-1) \},
%\end{align}
%becomes the subgroup of $G$, $H_N = Z_N \subset G$, and then $G$ can be decomposed as
%\begin{align}
%G \simeq G_0 \rtimes Z_N. \label{eq:G0ZN}
%\end{align}
%because of $G=G_0Z_N$ and $G_0 \cap Z_N = \{e\}$. (See also Appendix~\ref{app:semidirect}).
%In this case, the anomaly-free and anomalous parts of $G$ can be separated.
In more general, if there exists $\exists g \in G$, which satisfies $N(g)=N$ and ${\rm gcd}(Q(g),N(g))=1$, $G$ can be expressed as Eq.~(\ref{eq:G0ZN}) since this $g$ generates $Z_N$ subgroup of $G$ and it satisfies $G=G_0Z_N$ and $G_0 \cap Z_N = \{e\}$.
In particular, when $N$ is a prime number, they are automatically satisfied.

Let us summarize the important points in this section again.
\begin{itemize}
\item Generally, the anomaly-free subset of $G$, $G_0$, becomes a normal subgroup of $G$, $G_0 \triangleleft G$, and then the anomalous part becomes 
$G/G_0 \simeq Z_N$, where ${\rm det}\rho(g) \ (\forall g \in G)$ can be expressed as ${\rm det}\rho(g)=e^{2\pi i Q'(g)/N}$.
\item In particular, if there exists $\exists g \in G$ which satisfies $N(g)=N$ and ${\rm gcd}(Q(g),N(g))=1$, $G$ can be expressed as $G \simeq G_0 \rtimes Z_N$.
\end{itemize}

% ------------------------------------------------------ %
% ------------------------------------------------------ %
% ------------------------------------------------------ %
% ------------------------------------------------------ %

\section{Derived subgroup}
\label{sec:group}

In the previous section, by use of the representation of $\forall g \in G$, $\rho(g)$, we have shown  $G/G_0 \simeq Z_N$  generally.
However, even without use of explicit representations, this result is still valid although we do not know the explicit number $N$ 
from the beginning.
In this section, we study the detail structure of $G_0$ and $G/G_0 \simeq Z_N$ from the structure of $G$.
We note that the analysis in this section can be applied for any representations although we often consider a specific case.

First of all, we introduce the derived subgroup $D(G)$ of $G$, 
\begin{align}
D(G) \equiv <xyx^{-1}y^{-1} \in G | x, y \in G >, \label{eq:D(G)}
\end{align}
which is also called the commutator subgroup.
The derived subgroup $D(G)$ is a normal subgroup $G$, $D(G) \triangleleft G$, which can be shown as
\begin{align}
g(xyx^{-1}y^{-1})g^{-1}=(gxg^{-1})(gyg^{-1})(gxg^{-1})^{-1}(gyg^{-1})^{-1} \in D(G), \notag
\end{align}
for $\forall xyx^{-1}y^{-1} \in D(G)$ and $\forall g \in G$.
The quotient $G/D(G)$ is Abelian, which means that any cosets $D(G)X$ and $D(G)Y$ with $X,Y \notin D(G)$ satisfy $(D(G)X)(D(G)Y)=(D(G)Y)(D(G)X)$, because of $XYX^{-1}Y^{-1} \in D(G)$.
Indeed, the derived subgroup $D(G)$ is smallest among normal subgroups $K_G$ of $G$ such that 
$G/K_G$ is Abelian\footnote{If $G$ itself is an Abelian group, $D(G)=\{e\}$ is satisfied.}.
Thus, we can find that
\begin{align}
G_0 \supseteq D(G) \label{eq:G0D(G)},
\end{align}
because of $G/G_0 \simeq Z_N$.
We can actually check it as
\begin{align}
{\rm det} \rho(xyx^{-1}y^{-1}) = {\rm det}[\rho(x)\rho(y)\rho(x)^{-1}\rho(y)^{-1}] = 1.
\end{align}
%Furthermore,  for $xyx^{-1}y^{-1} \in D(G)$, we find 
%\begin{align}
%{\rm det} \rho(xyx^{-1}y^{-1}) = {\rm det}[\rho(x)\rho(y)\rho(x)^{-1}\rho(y)^{-1}] = 1.
%\end{align}
%That implies 
%\begin{align}
%G_0 \supseteq D(G).
%\end{align}
Therefore, the derived subgroup $D(G)$ of $G$ is always anomaly free.
The whole anomaly-free subgroup $G_0$ is either the same as $D(G)$ or larger than $D(G)$. 
We study how large $G_0$ is compared with $D(G)$ in what follows.

The order of $G/D(G)$ can be written as a product of prime numbers $p_i$, 
$|G/D(G)|=\prod_{i=1}^{r} p_i^{A_i}=p_1^{A_1} \cdots p_r^{A_r}$.
Then, by the fundamental structure theorem\footnote{See also Appendix \ref{app:Abelian}.} of finite Abelian group, 
$G/D(G)$ can be generally expressed as
\begin{align}
G/D(G) \simeq
(Z_{p_1^{a_{1,1}}}
%\times Z_{p_1^{a_{1,2}}}
\times \cdots \times Z_{p_1^{a_{1,n_1}}}) \times
%Z_{p_2^{a_{2,1}}} \times Z_{p_2^{a_{2,2}}} \times \cdots \times Z_{p_2^{a_{2,n_2}}}
\cdots \times
(Z_{p_r^{a_{r,1}}}
%\times Z_{p_r^{a_{r,2}}}
\times \cdots \times Z_{p_r^{a_{r,n_r}}}),
\label{eq:generalGD(G)}
\end{align}
where $a_{i,j}$ satisfy
\begin{align}
A_i = \sum_{j=1}^{n_i} a_{i,j}, \quad a_{i,j} \geq a_{i,j+1}. \notag
\end{align}
On the other hand, $G/G_0$ is isomorphic to a single cyclic group $Z_N$.
Thus, the structure of the Abelian group $G/D(G)$ in Eq.~(\ref{eq:generalGD(G)}) is important to 
study how large $G_0$ is compared with $D(G)$.

Now, let us classify our particle theories by the determinant of the representation (including reducible representation) of the element $X_{i,j}$ of $Z_{p_i^{a_{i,j}}}$ in the Abelian group $G/D(G)$~\footnote{Here, $Z_{p_i^{a_{i,j}}}$ is a subgroup of the Abelian group $G/D(G)$ but not a subgroup of $G$ in general. Then, the element $X_{i,j} \in Z_{p_i^{a_{i,j}}}$ corresponds to an element of the coset $D(G)X_{i,j}$ in $G$. However, since it is satisfied that ${\rm det}\rho(g_D)=1\ \forall g_D \in D(G)$, the determinant ${\rm det}\rho(X_{i,j})$ reflects on also $G$ as the same way.}, ${\rm det}\rho(X_{i,j})$.
%Now, let us classify the cases by ${\rm det}\rho(X_{i,j})$, where $X_{i,j}$ is an element of $Z_{p_i^{a_{i,j}}}$ which is a subgroup of the Abelian group $G/D(G)$.
%\textcolor{red}{
%Note that this classification depend on $\rho(X)$ in $G/D(G)$ appearing in our particle theories, that is, the representation of $X$ including 
%reducible representations.}
\begin{itemize}
\item[(i)]
%Suppose that 
We consider the theory in which
the element $\forall X_{i,j} \in Z_{p_i^{a_{i,j}}}$ for $\forall i,j$ satisfies
\begin{align}
{\rm det}\rho(X_{i,j})=1.
\end{align}
In this case, we can easily find that
\begin{align}
G_0=G.
\end{align}
\item[(ii)]
%Suppose that 
We consider the theory in which
the element $\exists X_{i,j} \in Z_{p_i^{a_{i,j}}}-\{e\}$ for $\exists i,j$ satisfies
\begin{align}
{\rm det}\rho(X_{i,j})=1,
\end{align}
while any other element $\forall X'_{i,j} \in Z_{p_i^{a_{i,j}}}-\{e,X_{i,j}\}$ for $\forall i,j$ satisfies
\begin{align}
{\rm det}\rho(X'_{i,j}) \neq 1.
\end{align}
In this case, we can find that
\begin{align}
G \supset G_0 \supset D(G).
\end{align}
Although we do not discuss the detail of this class in this paper since there are several patterns of this type in some discrete groups $G$, one can also consider this class similarly as the following analysis.
\item[(iii)]
%Suppose that 
We consider the theory in which
the element $\forall X_{i,j} \in Z_{p_i^{a_{i,j}}}-\{e\}$ for $\forall i,j$ satisfies
\begin{align}
{\rm det}\rho(X_{i,j}) \neq 1. \label{eq:not1}
\end{align}
%However, 
If there exists $a_{i,2} \neq 0$ for $\exists i$, $p_i^{a_{i,2}}-1$ numbers of combination elements $X_{i,1}^{m_1}X_{i,2}^{m_2}$ satisfy 
\begin{align}
{\rm det}\rho(X_{i,1}^{m_1}X_{i,2}^{m_2})=1,
\end{align}
where $m_1$ and $m_2$ satisfy the following relation;
\begin{align}
\begin{array}{ll}
&Q(X_{i,1}^{m_1}X_{i,2}^{m_2}) = m_1 Q(X_{i,1}) + m_2 Q(X_{i,2}) p_i^{a_{i,1}-a_{i,2}} \equiv 0 \quad ({\rm mod}\ p_i^{a_{i,1}}), \\
\Leftrightarrow
&m_1 Q(X_{i,1}) = p_i^{a_{i,1}-a_{i,2}} n, \quad  m_2 Q(X_{i,2}) = p_i^{a_{i,2}} - n, \quad \forall n \in \mathbb{Z}/p_i^{a_{i,2}}\mathbb{Z}-\{0\}.
\end{array}
\end{align}
Then, those $X_{i,1}^{m_1}X_{i,2}^{m_2}$ as well as $e$ are also included in $G_0$, which means
\begin{align}
G \supset G_0 \supset D(G),
\end{align}
and also they construct $Z_{p_i^{a_{i,2}}}$ subgroup of $G_0/D(G)$.
Note that, in general, if there are $Z_{N_1}$ and $Z_{N_2}$ symmetries in $G/D(G)$, where $N_1$ and $N_2$ are not coprime to each other, ${\rm gcd}(N_1,N_2)-1$ numbers of elements $X_{1}^{m_1}X_{2}^{m_2}\ (X_1 \in Z_{N_1}, X_2 \in Z_{N_2})$ as well as $e$ satisfy ${\rm det}\rho(X_1X_2)=1$ and then they construct $Z_{{\rm gcd}(N_1,N_2)}$ subgroup of $G_0/D(G)$.
In a similar way, we can find that
\begin{align}
G_0/D(G) \simeq
(Z_{p_1^{a_{1,2}}}
%\times Z_{p_1^{a_{1,2}}}
\times \cdots \times Z_{p_1^{a_{1,n_1}}}) \times
%Z_{p_2^{a_{2,1}}} \times Z_{p_2^{a_{2,2}}} \times \cdots \times Z_{p_2^{a_{2,n_2}}}
\cdots \times
(Z_{p_r^{a_{r,2}}}
%\times Z_{p_r^{a_{r,2}}}
\times \cdots \times Z_{p_r^{a_{r,n_r}}}).
\label{eq:generalG0D(G)}
\end{align}
Then, by using the isomorphism theorem $3$ in Appendix~\ref{app:isomorphism} and Eqs.~(\ref{eq:generalGD(G)}) and (\ref{eq:generalG0D(G)}), we can obtain
\begin{align}
Z_N \simeq G/G_0 \simeq (G/D(G))/(G_0/D(G)) \simeq Z_{(p_1^{a_{1,1}} \cdots p_r^{a_{r,1}})}. \label{eq:GG0iii}
\end{align}
Indeed, when Eq.~(\ref{eq:not1}) is satisfied, the determinant ${\rm det}\rho(g)$ for $\forall g \in G$ can be expressed as ${\rm det}\rho(g)=e^{2\pi iQ'(g)/(p_1^{a_{1,1}} \cdots p_r^{a_{r,1}})}$, which means that we get Eq.~(\ref{eq:GG0iii}).
Note that $N= p_1^{a_{1,1}} \cdots p_r^{a_{r,1}} = \prod_{i=1}^{r} p_i^{a_{i,1}}$ is the least common multiple of orders of each $Z_{p_i^{a_{i,j}}}$ in $G/D(G)$ and it becomes the maximum order of the anomalous $G/G_0 \simeq Z_N$.
In other words, the maximum order of the anomalous $G/G_0 \simeq Z_N$ can be found by $G/D(G)$, which is determined by $G$.

In particular, if and only if it is also satisfied that $A_i=a_{i,1}\ (a_{i,2}=0)$ for all $i$, any element $X$ in the Abelian group $G/D(G)$ leads to
\begin{align}
{\rm det}\rho(X) \neq 1,
\end{align}
which means that
\begin{align}
G_0=D(G),
\end{align}
and then, by using Eq.~(\ref{eq:mn}) in Appendix~\ref{app:Abelian}, we can obtain
\begin{align}
Z_N \simeq G/G_0 = G/D(G) \simeq Z_{(p_1^{A_1} \cdots p_r^{A_r})}.
\end{align}
Therefore, $D(G)$ gives us an important clue to obtain information about $G_0$ and $G/G_0 \simeq Z_N$.
In the following analysis, we mainly discuss this class (iii) that Eq.~(\ref{eq:not1}) is satisfied.
\end{itemize}

% ------------------------------------------------------ %
% ------------------------------------------------------ %

\subsection{Various examples of $G$}
\label{subsec:example}

Now, in this subsection, let us see the detail structure of $G_0$ and $G/G_0 \simeq Z_N$ from the structure of $D(G)$ and $G/D(G)$ through specific examples.

First, when $G$ is a perfect group, defined as a group satisfying $D(G)=G$, obviously we can find that $G = G_0 = D(G)$.
In particular, non-Abelian simple groups such as $A_n\ (n \geq 5)$ and $PSL(2,Z_p)\ (p \neq 2,3,\ p \in \mathbb{P})$ are the simplest examples of perfect groups.
That is, the whole symmetry $G$ is always anomaly free.
(See also Ref.~\cite{Chen:2015aba}.)
Here, we mention about a simple group.
The simple group is defined as the group $G$, such that it does not have any normal subgroups but $\{e\}$ and $G$ itself. 
Then, since $G/D(G)$ is Abelian, $D(G)$ must be equal to $G$ itself for a non-Abelian simple group.

On the other hand, an Abelian simple group, which is just isomorphic to $Z_p\ (p \in \mathbb{P})$, is not a perfect group since 
$D(G)=\{e\}$.
In this case, the flavor model corresponds to either class (i), $G_0=G \simeq Z_p$, or the class (iii), $G_0=D(G)=\{e\}$, $G/G_0 \simeq Z_N=Z_p$.
The group $G = A_3 \simeq Z_3$ is an example of this case.

Next, let us consider the group $G$, which can be decomposed by a semidirect product,~i.e. $G \simeq K_G \rtimes G^{(1)}$.
We discuss this in the following four steps.

\begin{step}[$G \simeq Z_A \rtimes Z_B$]
Let us start with the simplest group, $G \simeq Z_A \rtimes Z_B\ (G/Z_A \simeq Z_B)$, where $A \geq 3$.
Then, it is found that $D(G) \subseteq Z_A$.
When we take $\alpha \in Z_A$ and $\beta \in Z_B$, 
they satisfy the following algebraic relations:
\begin{align}
\alpha^A = \beta^B = e
\end{align}
and also 
\begin{align}
&\beta \alpha \beta^{-1} = \alpha^m \in Z_A, \quad (\beta \alpha \beta^{-1} \alpha^{-1} = \alpha^{m-1} \in D(G)), \quad m \in \mathbb{Z}/A\mathbb{Z} - \{0,1\}, \label{eq:semidircond1ZAZB} \\
\Rightarrow \ 
&\beta^b \alpha^a \beta^{-b} = a^{am^b}, \quad (\beta^b \alpha^a \beta^{-b} \alpha^{-a} = a^{a(m^b-1)} \in D(G)),
\quad a \in \mathbb{Z}/A\mathbb{Z}, \ b \in \mathbb{Z}/B\mathbb{Z}, \label{eq:semidircond2ZAZB}
\end{align}
where $m$ satisfies\footnote{If $m=1$, the group$G$ reduces to $G \simeq Z_A \times Z_B$. This case is the specific case in the above general analysis.} the following conditions,
\begin{align}
\begin{array}{l}
(m^b - 1) = (m-1)(\sum_{r=0}^{b-1}m^{r}) \not\equiv 0 \quad ({\rm mod}\ A) \quad {\rm for}\ \forall b \\
(m^B - 1) = (m-1)(\sum_{r=0}^{B-1}m^{r}) \equiv 0 \quad ({\rm mod}\ A)
\end{array}. \label{eq:constm}
\end{align}
From Eq,~(\ref{eq:semidircond2ZAZB}), we find that
\begin{align}
D(G) = \{ \alpha^{a'{\rm gcd}(m-1,A)} | a' \in \mathbb{Z}/(A/{\rm gcd}(m-1,A))\mathbb{Z} \} = Z_{A/{\rm gcd}(m-1,A)} \subseteq G_0, \label{eq:D(G)ZAZB}
\end{align}
while $D(G)=\{e\}$ in the case of $G \simeq Z_A$.
Then, we find that
\begin{align}
G/D(G) \simeq Z_{{\rm gcd}(m-1,A)} \times Z_B, \label{eq:GD(G)ZAZB}
\end{align}
since cosets $D(G)\alpha$ and $D(G)\beta$ satisfy 
\begin{align}
\begin{array}{l}
(D(G)\alpha)^{{\rm gcd}(m-1,A)}=(D(G)\beta)^B=D(G), \\
(D(G)\beta)(D(G)\alpha)(D(G)\beta)^{-1}=D(G)(\beta\alpha\beta^{-1}\alpha^{-1})\alpha=D(G)\alpha.
\end{array}
\notag
\end{align}

The above result can also be understood in terms of $Z_A$ charge constraint.
Let us assume that the chiral fermions are $Z_A$ eigenstates, $\rho(\alpha)_{jk}=e^{2\pi iq_j/A}\delta_{jk}$, where $q_j$ is the $Z_A$ charge of $j$ th component.
Eq.~(\ref{eq:semidircond1ZAZB}) shows that there exists a state with charge $mq_j$ and 
$\beta$ transforms the $j$ th component to the state with $mq_j$.
Then, when we consider a fundamental irreducible representation, it becomes $B$-dimensional representation with $Z_A$ charge $ ^t(q_1,q_2,q_3,...,q_B)$ $=$ $ ^t(q, mq, m^2q,...,m^{B-1}q)$.
It means that $Z_A$ charges in a multiplet are constrained by the semidirect product by $Z_B$, while there is no constraint for $Z_B$ charges.
Thus, we obtain
\begin{align}
{\rm det}\rho(\alpha) = e^{2\pi i\sum_{r=0}^{B-1}m^rq/A} = e^{2\pi iqn/{\rm gcd}(m-1,A)},
\end{align}
which means Eqs.~(\ref{eq:D(G)ZAZB}) and (\ref{eq:GD(G)ZAZB}).

Now, as discussed in the above general analysis, let us study the class (iii).
Hereafter, we use $A'\equiv {\rm gcd}(m-1,A)$.
Then, the elements, $\alpha^x\beta^y$, which satisfy
\begin{align}
&Q(\alpha^x)B/{\rm gcd}(A',B) + Q(\beta^y)A'/{\rm gcd}(A',B) \equiv 0 \quad ({\rm mod}\ {\rm lcm}(A',B)) \label{eq:QxQy} \\
\Leftrightarrow \quad
&xQ(\alpha) = ({\rm gcd}(A',B) - s) A'/{\rm gcd}(A',B), \ yQ(\beta) = s B/{\rm gcd}(A',B), \ s \in \mathbb{Z}/{\rm gcd}(A',B)\mathbb{Z}, \notag
\end{align}
also satisfy ${\rm det}\rho(\alpha^x\beta^y)=1$, that is, the elements of ${\rm gcd}(A',B)$ numbers of cosets $D(G)\alpha^x\beta^y$ (including $D(G)\alpha^{A'}=D(G)$) become the elements of $G_0$.
In addition, since these cosets satisfy $(D(G)\alpha^x\beta^y)^{{\rm gcd}(A',B)}=D(G)$, they construct 
the (normal) subgroup of $G/D(G)$,
\begin{align}
G_0/D(G) \simeq Z_{{\rm gcd}(A',B)}.
\end{align}
%(normal) subgroup of $G/D(G)$.
On the other hand, the generators of $G_0$ are $\alpha^{A'}$ and $\alpha^x\beta^y$, where $x$, $y$ satisfy the above conditions in Eq.~(\ref{eq:QxQy}), and they satisfy
\begin{align}
(\alpha^{A'})^{A/A'}=e, \ 
(\alpha^x\beta^y)^{{\rm gcd}(A',B)} = (\alpha^{A'})^{k=\frac{x(m^{y{\rm gcd}(A',B)}-1)}{A'(m^y-1)}},\ 
(\alpha^x\beta^y)(\alpha^{A'})(\alpha^x\beta^y)^{-1}=(\alpha^{A'})^{m^y}. 
\label{eq:alphabetacondition}
\end{align}
In particular, if $k \equiv 0\ ({\rm mod}\ A/A')$, which means $\alpha^x\beta^y$ actually generates this $Z_{{\rm gcd}(A',B)}$ subgroup of $G_0$, $G_0$ can be written as $G_0 \simeq D(G) \rtimes Z_{{\rm gcd}(A',B)} = Z_{A/A'} \rtimes Z_{{\rm gcd}(A',B)}$.
Then, by using the isomorphism theorem $3$ in Appendix~\ref{app:isomorphism}, we can obtain
\begin{align}
Z_N \simeq G/G_0 \simeq (G/D(G))/(G_0/D(G)) \simeq (Z_{A'} \times Z_B)/Z_{{\rm gcd}(A',B)} \simeq Z_{{\rm lcm}(A',B)},
\end{align}
where we also use Eq.~(\ref{eq:gcdlcm}) in Appendix~\ref{app:Abelian} for Eq.~(\ref{eq:GD(G)ZAZB}),
\begin{align}
G/D(G) \simeq Z_{A'} \times Z_{B} \simeq Z_{{\rm gcd}(A',B)} \times Z_{{\rm lcm}(A,B)}. \label{eq:GD(G)gcdlcm}
\end{align}
Similar to Eq.~(\ref{eq:QxQy}), there exists $\alpha^{x'}\beta^{y'}$ with $\exists (x',y')$, which satisfies
\begin{align}
Q(\alpha^{z'})B/{\rm gcd}(A',B) + Q(\beta^{y'})A'/{\rm gcd}(A',B) \equiv 1 \quad ({\rm mod}\ {\rm lcm}(A',B)),
\end{align}
since $A'/{\rm gcd}(A',B)$ and $B/{\rm gcd}(A',B)$ are coprime to each other.
It means that this element $\alpha^{x'}\beta^{y'}$ corresponds to $g_1$ in the previous section.
In addition, the coset $D(G)\alpha^{x'}\beta^{y'} \subset G_0\alpha^{x'}\beta^{y'}$ satisfies $(D(G)\alpha^{x'}\beta^{y'})^{{\rm lcm}(A',B)}=D(G) \subset G_0$.
Then, similar to Eq.~(\ref{eq:alphabetacondition}), 
this element generally satisfies 
\begin{align}
\begin{array}{l}
(\alpha^{x'}\beta^{y'})^{{\rm lcm}(A',B)} = (\alpha^{A'})^{k'=\frac{x'(m^{y'{\rm lcm}(A',B)}-1)}{A'(m^{y'}-1)}}, \\
(\alpha^{x'}\beta^{y'})(\alpha^{A'})(\alpha^{x'}\beta^{y'})^{-1}=(\alpha^{A'})^{m^{y'}}, \\
(\alpha^{x'}\beta^{y'})(\alpha^x\beta^y)(\alpha^{x'}\beta^{y'})^{-1}=(\alpha^{A'})^{\ell'=[x(m^{y'}-1)-x'(m^y-1)]/A'}(\alpha^x\beta^y).
\end{array}
\end{align}
%are generally satisfied, and 
If $k' \equiv 0\ ({\rm mod}\ A/A')$, which means $\alpha^{x'}\beta^{y'}$ actually generates this $Z_{{\rm lcm}(A',B)}$ subgroup of $G$, $G$ can be similarly written as $G \simeq G_0 \rtimes Z_{{\rm lcm}(A',B)}$.
In addition, if $\ell' \equiv 0\ ({\rm mod}\ A/A')$, the element $\alpha^{x'}\beta^{y'}$ commutes $\alpha^x\beta^y$.
Thus, if $k, k', \ell'$ are multiples of $A/A'$, which means $Z_{{\rm gcd}(A',B)} \times Z_{{\rm lcm}(A,B)}$ in Eq.~(\ref{eq:GD(G)gcdlcm}) (as well as each of $Z_{{\rm gcd}(A',B)}$ and $Z_{{\rm lcm}(A,B)}$) is actually a subgroup of $G$, $G$ can be written as
\begin{align}
\begin{array}{rl}
G
&\simeq D(G) \rtimes ( Z_{{\rm gcd}(A',B)} \times Z_{{\rm lcm}(A,B)} ) \simeq Z_{A/A'} \rtimes (Z_{A'} \times Z_B) \\
&\simeq (D(G) \rtimes Z_{{\rm gcd}(A',B)}) \rtimes Z_{{\rm lcm}(A,B)} \\
&\simeq G_0 \rtimes Z_{{\rm lcm}(A,B)}, \label{eq:GD(G)G0}
\end{array}
\end{align}
where the second line can be understood by Eqs.~(\ref{eq:decomposiG1}) and (\ref{eq:decomposiK}) in Appendix~\ref{app:semidirect}.

Now, let us see examples.
First, if $A=p \in \mathbb{P}$, 
we obtain $A'=1$.
Then,  it is found $D(G)=Z_A \subseteq G_0$.
When we consider the class (iii), 
%If this does not  the class (i), 
we obtain $G_0=D(G)=Z_A$ and $G/G_0 \simeq Z_N = Z_B$.

Second, let us study $G = D_A \simeq Z_A \rtimes Z_2$.
In this case, we obtain $m=A-1$ from Eq.~(\ref{eq:constm}), and then we can find that
\begin{align}
&D(G)
= \left\{
\begin{array}{ll}
< \tilde{\alpha} | \tilde{\alpha}^{A/2} = e > = Z_{A/2} & (A \in 2\mathbb{Z}) \\
< \alpha | \alpha^A = e > = Z_A & (A \in 2\mathbb{Z}+1)
\end{array}
\right., \\
&G/D(G)
\simeq \left\{
\begin{array}{ll}
Z_{2} \times Z_2 = < (d_{\alpha},d_{\beta}) | d_{\alpha}^2=d_{\beta}^2=d_{e}, d_{\alpha}d_{\beta}=d_{\beta}d_{\alpha}=d_{\alpha\beta} > & (A \in 2\mathbb{Z}) \\
Z_2 = < d_{\beta} | d_{\beta}^2=d_{e} > & (A \in 2\mathbb{Z}+1)
\end{array}
\right.,
\end{align}
%\begin{align}
%D(D_A) = \left\{
%\begin{array}{ll}
%Z_{A/2} & (A \in 2\mathbb{Z}) \\
%Z_A & (A \in 2\mathbb{Z}+1)
%\end{array}
%\right., \quad
%D_A/D(D_A) \simeq \left\{
%\begin{array}{ll}
%Z_{2} \times Z_2 & (A \in 2\mathbb{Z}) \\
%Z_2 & (A \in 2\mathbb{Z}+1)
%\end{array}
%\right.,
%\end{align}
are satisfied,
%\footnote{$Q_A$ group is similar to $D_A$ with $A \in 2\mathbb{Z}$. The difference is $\beta^2=\alpha^{A/2}$ instead of $\beta^2=e$. It means the determinant of $\beta$ depends on that of $\alpha$, although $D(Q_A)=D(D_A)$. In the case of $A \in 4\mathbb{Z}$, since $\alpha^{A/2}$ as well as $e$ are included in $D(Q_A)$, the situation in this analysis is same as $D_A$. However, in the case of $A \in 2(2\mathbb{Z}+1)$, since $\alpha^{A/2} \notin D(Q_A)$, $\beta$ becomes anomalous $Z_4$ generator, in general. This is different from $D_A$.},
where $\tilde{\alpha} \equiv \alpha^2$, $d_{X} \equiv D(G)X$.
Then, %when we consider the same situation as Eq.~(\ref{eq:min}), 
in the class (iii), 
we can obtain
\begin{align}
&G_0 = \left\{
\begin{array}{ll}
< \tilde{\alpha}, \tilde{\beta} | \tilde{\alpha}^{A/2}=\tilde{\beta}^2=e, \tilde{\beta}\tilde{\alpha}\tilde{\beta}^{-1}=\tilde{\alpha}^{-1} > = Z_{A/2} \rtimes Z_2 \simeq D_{A/2} & (A \in 2\mathbb{Z}) \\
< \alpha | \alpha^A=e > = Z_A & (A \in 2\mathbb{Z}+1)
\end{array}
\right., \\
&G/G_0 \simeq Z_N = Z_2 = < g_{\beta} | g_{\beta}^2=g_{e} >, \quad \forall A,
\end{align}
%\begin{align}
%G_0 = \left\{
%\begin{array}{ll}
%Z_{A/2} \rtimes Z_2 \simeq D_{A/2} & (A \in 2\mathbb{Z}) \\
%Z_A & (A \in 2\mathbb{Z}+1)
%\end{array}
%\right., \quad
%D_A/G_0 \simeq Z_N = Z_2, \quad \forall A,
%\end{align}
where $\tilde{\beta} \equiv \alpha\beta \ (x=y=1)$, $g_{\beta} \equiv G_0\beta \ (x'=0, y'=1)$.
Then, since $\beta^2=\tilde{\beta}^2=e$,
we can write
\begin{align}
\begin{array}{rl}
D_A=G
&\simeq G_0 \rtimes Z_2 \\
&\simeq \left\{
\begin{array}{ll}
(D(G) \rtimes Z_2) \rtimes Z_2 &  (A \in 2\mathbb{Z}) \\
D(G) \rtimes Z_2 &  (A \in 2\mathbb{Z}+1)
\end{array}
\right. \\
&\simeq \left\{
\begin{array}{ll}
D_{A/2} \rtimes Z_2 &  (A \in 2\mathbb{Z}) \\
Z_A \rtimes Z_2 &  (A \in 2\mathbb{Z}+1)
\end{array}
\right..
\end{array}
\end{align}
Here, we comment on the case with $A=2(2m-1)\ (m \in \mathbb{Z})$ in particular.
In this case, since the coset relation $g_{\beta}=g_{\alpha^{A/2}}\ (x'=A/2,y'=0)$ is satisfied and then $k, k', \ell'$ are multiples of $A/A'=A/2$, Eq.~(\ref{eq:GD(G)G0}) is satisfied;
\begin{align}
D_A=G
&\simeq D(G) \rtimes (Z_2 \times Z_2) \simeq  Z_{A/2} \rtimes (Z_2 \times Z_2) \notag \\
&\simeq G_0 \times Z_2 \simeq D_{A/2} \times Z_2. \notag
\end{align}
For example, the group $G= S_3 \simeq D_3 \simeq A_3 \rtimes Z_2$ is included in this case.
Then, we can find that
the $S_3$ flavor model corresponds to either the class (iii), $G_0=D(G)=Z_3 \simeq A_3$, $G/G_0 \simeq Z_N=Z_2$, or the class (i), $G_0=G=S_3$.
It depends on representations including reducible ones.
Indeed, $S_3$ have three irreducible representations, ${\bf 1}$, ${\bf 1'}$, and ${\bf 2}$, which have ${\rm det}\rho_{{\bf 1}}(\beta)=1$, ${\rm det}\rho_{{\bf 1'}}(\beta)={\rm det}\rho_{{\bf 2}}(\beta)=-1$~\cite{Ishimori:2010au,Ishimori:2012zz}.
%with 
%$\rho({\bf 1})=1$, $\rho({\bf 1'})=\rho({\bf 2})=-1$ \cite{Ishimori:2010au,Ishimori:2012zz}.
The whole $S_3$ symmetry is anomaly free in flavor models including even numbers of ${\bf 1'}$ and ${\bf 2}$.
Otherwise, the $Z_2$ subsymmetry can be anomalous.

We comment on $G=Q_A$ case.
$Q_A$ group is similar to $D_A$ with $A \in 2\mathbb{Z}$.
The difference is $\beta^2=\alpha^{A/2}$ instead of $\beta^2=e$.
It means that the determinant of the representation of $\beta$ depends on that of $\alpha$, although $D(Q_A)=D(D_A)$.
In the case of $A \in 4\mathbb{Z}$, since $\alpha^{A/2}$ as well as $e$ are included in $D(Q_A)$, the analysis of $Q_A$ is same as that of $D_A$.
However, in the case of $A \in 2(2\mathbb{Z}+1)$, since $\alpha^{A/2} \notin D(Q_A)$, $\beta$ becomes anomalous $Z_4$ generator in general.
This is different from $D_A$.

Third, let us consider $G=T_{p^k} \simeq Z_{p^k} \rtimes Z_3$, where $p \neq 3$ and $p \in \mathbb{P}$.
If ${\rm gcd}(m-1,p^k) \neq 1$, it should be satisfied that $m=p^{\ell}+1$, where $\ell \in \mathbb{Z}/k\mathbb{Z}-\{0\}$.
That requires 
\begin{align}
m^3-1 = p^{\ell} (p^{2\ell}+3p^{\ell}+3) \equiv 0 \ ({\rm mod}\ p^k) \notag \\
\Rightarrow \ 
p^{2\ell}+3p^{\ell}+3 = p^{k-\ell}x, \quad (x \in \mathbb{Z}) \notag \\
3 = p^{k-\ell}x - p^{2\ell} - 3p^{\ell} = p^{\ell'}y .
\end{align}
However, it cannot be satisfied if $p\neq3$.
Thus, we obtain ${\rm gcd}(m-1,p^k) = 1$ and then $D(G)=Z_{p^k} \subseteq G_0$.
Therefore, the $T_{p^k}$ flavor model corresponds to either the class (iii), $G_0=D(G)=Z_{p^k}$, $G/G_0 \simeq Z_N=Z_3$, or the class (i), $G_0=G=T_{p^k}$.
In more general, when we discuss $G \simeq Z_{p^k} \rtimes Z_B$ and ${\rm gcd}(p,B)=1$, 
we obtain ${\rm gcd}(m-1,p^k)=1$ and then we find $D(G)=Z_{p^k} \subseteq G_0$.
\end{step}

\begin{step}[$G \simeq (Z_A \times Z'_A) \rtimes Z_B$]
Next, let us consider more complicate case, $G \simeq (Z_A \times Z'_{A}) \rtimes Z_B$.
%However, it is difficult to discuss generic $Z_B$.
%combination of $Z_A$ and $Z_A'$.
%Then, let us see the following examples.
%
First, let us consider $G=\Sigma(2n^2) \simeq (Z_n \times Z'_n) \rtimes Z_2$.
When we take $\alpha \in Z_n$, $\alpha' \in Z'_n$, and $\beta \in Z_2$, 
they satisfy the following algebraic relations:
\begin{align}
\alpha^n = \alpha^{\prime n} = \beta^2 = e,
\end{align}
and also
\begin{align}
&\alpha' \alpha \alpha^{\prime -1} = \alpha \in Z_n, \quad (\alpha \alpha' \alpha^{-1} = \alpha' \in Z'_n) \\
&\beta \alpha \beta^{-1} = \alpha^{m_1} \alpha^{\prime m_2} \in Z_n \times Z'_n, \quad \beta \alpha^{-1} \beta^{-1} = \alpha^{m_3} \alpha^{\prime m_4} \in Z_n \times Z'_n. \label{eq:semidircond1ZnZnZ2}
\end{align}
Here, $m_i\ (i=1,2,3,4)$ can be determined by the constraints, $\beta^2 \alpha \beta^{-2}=\alpha$, $\beta^2 \alpha' \beta^{-2}=\alpha'$,
%\begin{align}
%\alpha
%&= \beta^2 \alpha \beta^{-2} \notag \\
%&= \beta \alpha^{m_1} \beta^{-1} \beta \alpha'^{m_2} \beta^{-1} \notag \\
%&= \alpha^{m_1^2} \alpha'^{m_1m_2} \alpha^{m_2m_3} \alpha'^{m_2m_4} \notag \\
%&= \alpha^{m_1^2+m_2m_3} \alpha'^{m_2(m_1+m_4)}, \\
%\alpha'
%&= \beta^2 \alpha' \beta^{-2} \notag \\
%&= \beta \alpha^{m_3} \beta^{-1} \beta \alpha'^{m_4} \beta^{-1} \notag \\
%&= \alpha^{m_3m_1} \alpha'^{m_3m_2} \alpha^{m_4m_3} \alpha'^{m_4^2} \notag \\
%&= \alpha^{m_3(m_1+m_4)} \alpha'^{m_4^2+m_2m_3},
%\end{align}
and then we obtain $m_1=m_4=0,\ m_2=m_3=1$.
Thus, Eq.~(\ref{eq:semidircond1ZnZnZ2}) can be rewritten as
\begin{align}
\beta \alpha \beta^{-1} = \alpha', \quad \beta \alpha' \beta^{-1} = \alpha. \label{eq:semidircond2ZnZnZ2}
\end{align}
Then, we can find that
\begin{align}
&D(G) = < \tilde{\alpha} | \tilde{\alpha}^n=e > = Z_n, \label{eq:D(G)ZnZnZ2} \\
&G/D(G) \simeq Z_n \times Z_2 = < (d_{\tilde{\alpha}'}, d_{\beta}) | d_{\tilde{\alpha}'}^n=d_{\beta}^2=d_e, d_{\tilde{\alpha}'}d_{\beta}=d_{\beta}d_{\tilde{\alpha}'}=d_{\alpha\beta} >, \label{eq:GD(G)ZnZnZ2}
\end{align}
where $\tilde{\alpha} \equiv \alpha \alpha^{\prime -1}$, $d_X \equiv D(G)X$.
%\begin{align}
%D(\Sigma(2n^2)) = < \tilde{\alpha} \equiv \alpha \alpha^{\prime -1} | \tilde{\alpha}^n=e > = Z_n, \quad \Sigma(2n^2)/D(\Sigma(2n^2)) \simeq Z_n \times Z_2. \label{DGDZnZnZ2}
%\end{align}

The above situation can also be understood in terms of $Z_n$ and $Z'_n$ charge constraints.
Let us assume that the chiral fermions are $Z_n$ and $Z'_n$ eigenstates, where $[Z_n, Z'_n]$ charges of the $j$ th component field are $[q_j, q'_j]$.
Eq.~(\ref{eq:semidircond2ZnZnZ2}) shows that there exists a state with charge $[q'_j, q_j]$ and the 
$\beta$ transforms the $j$ th component to the state with charge $[q'_j, q_j]$.
When we consider a fundamental irreducible representation, it becomes a doublet with charge $ ^t([q_1,q'_1], [q_2, q'_2])$ $=$ $ ^t([q, q'], [q', q])$, and then we obtain 
\begin{align}
{\rm det}\rho(\alpha) = {\rm det}\rho(\alpha') = e^{2\pi i (q+q')/n},
\end{align}
which means Eqs.~(\ref{eq:D(G)ZnZnZ2}) and (\ref{eq:GD(G)ZnZnZ2}).
Note that there is no constraint for $Z_2$ charges.

Similarly, in the class (iii), we can obtain
\begin{align}
&G_0 = \left\{
\begin{array}{ll}
< \tilde{\alpha}, \tilde{\beta} | \tilde{\alpha}^n=e,  \tilde{\beta}^2=\tilde{\alpha}^{n/2}, \tilde{\beta}\tilde{\alpha}\tilde{\beta}^{-1}=\tilde{\alpha}^{-1} > = Q_n & (n \in 2\mathbb{Z}) \\
< \tilde{\alpha} | \tilde{\alpha}^n=e > = Z_n & (n \in 2\mathbb{Z}+1)
\end{array}
\right., \\
&G/G_0 \simeq Z_N = \left\{
\begin{array}{ll}
Z_n = < g_{\alpha} | g_{\alpha}^n=g_e > & (n \in 2\mathbb{Z}) \\
Z_n \times Z_2 \simeq Z_{2n} = <g_{\tilde{\gamma}} | g_{\tilde{\gamma}}^{2n}=g_e > & (n \in 2\mathbb{Z}+1)
\end{array}
\right.,
\end{align}
where $\tilde{\beta} \equiv \alpha^{-n/2}\beta$, $\tilde{\gamma} \equiv \alpha^{-(n-1)/2}\beta$, $g_X \equiv G_0X$.
%\begin{align}
%\begin{array}{l}
%G_0 = \left\{
%\begin{array}{ll}
%< \tilde{\alpha}, \tilde{\beta} \equiv \alpha^{\prime n/2}\beta | \tilde{\alpha}^n=e,  \tilde{\beta}^2=\tilde{\alpha}^{n/2}, \tilde{\beta}\tilde{\alpha}\tilde{\beta}^{-1}=\tilde{\alpha}^{-1} > = Q_n & (n \in 2\mathbb{Z}) \\
%< \tilde{\alpha} | \tilde{\alpha}^n=e > = Z_n & (n \in 2\mathbb{Z}+1)
%\end{array}
%\right., \\
%\Sigma(2n^2)/G_0 \simeq Z_N = \left\{
%\begin{array}{ll}
%Z_n & (n \in 2\mathbb{Z}) \\
%Z_n \times Z_2 \simeq Z_{2n} & (n \in 2\mathbb{Z}+1)
%\end{array}
%\right.
%\end{array}
%\end{align}
We also introduce $\tilde{\alpha}' \equiv \tilde{\gamma}^2 = \tilde{\alpha}^{-(n-1)/2}\alpha'$ and $\tilde{\beta}' \equiv \tilde{\gamma}^n = \tilde{\alpha}^{-(n^2-1)/4}\beta$.
In particular, since $\alpha^n=\tilde{\gamma}^{2n}(=\tilde{\alpha}^{\prime n}=\tilde{\beta}^{\prime 2})=e$,
$G=\Sigma(2n^2)$ can be written as
\begin{align}
\begin{array}{rl}
\Sigma(2n^2)=G
&\simeq G_0 \rtimes Z_N \\
&\simeq \left\{
\begin{array}{ll}
Q_n \rtimes Z_n & (n \in 2\mathbb{Z}) \\
Z_n \rtimes Z_{2n} & (n \in 2\mathbb{Z}+1)
\end{array}
\right. \\
&\simeq \left\{
\begin{array}{ll}
Q_n \rtimes Z_n & (n \in 2\mathbb{Z}) \\
(Z_n \rtimes Z_{2}) \times Z_n \simeq D_n \times Z_n & (n \in 2\mathbb{Z}+1)
\end{array}
\right..
\end{array}
\end{align}
Here, the structure of the semidirect product in the case of $n \in 2\mathbb{Z}$ comes from $\alpha \tilde{\beta} \alpha^{-1} = \tilde{\alpha}\tilde{\beta}$, while the structures of semidirect and direct products in the case of $n \in 2\mathbb{Z}+1$ come from $\tilde{\beta}' \tilde{\alpha} \tilde{\beta}^{\prime -1} = \tilde{\alpha}^{-1}$, $\tilde{\alpha}' \tilde{\alpha} \tilde{\alpha}^{\prime -1} = \tilde{\alpha}$, $\tilde{\alpha}' \tilde{\beta}' \tilde{\alpha}^{\prime -1} = \tilde{\beta}'$.

This analysis can be easily applied to $G=\Sigma(3n^3) \simeq (Z_n \times Z'_n \times Z''_n) \rtimes Z_3$.
When we take $\alpha \in Z_n$, $\alpha' \in Z'_n$, $\alpha'' \in Z''_n$, and $\beta \in Z_3$, 
they satisfy the following algebraic relations:
\begin{align}
\begin{array}{l}
\alpha^n = \alpha^{\prime n} = \alpha^{\prime\prime n} = \beta^3 = e, \\
\alpha\alpha' = \alpha'\alpha, \ \alpha'\alpha'' = \alpha''\alpha', \ \alpha''\alpha = \alpha\alpha'', \\
\beta \alpha \beta^{-1} = \alpha', \ \beta \alpha' \beta^{-1} = \alpha'', \ \beta \alpha'' \beta^{-1} = \alpha.
\end{array}
\end{align}
Similarly, we can find that
\begin{align}
&D(G) = < \tilde{\alpha}, \tilde{\alpha}' | \tilde{\alpha}^n=\tilde{\alpha}^{\prime n}=e, \tilde{\alpha}\tilde{\alpha}'=\tilde{\alpha}'\tilde{\alpha} > = Z_n \times Z_n, \label{eq:D(G)ZnZnZnZ3} \\
&G/D(G) \simeq Z_n \times Z_3 = < (d_{\alpha}, d_{\beta}) | d_{\alpha}^n=d_{\beta}^3=d_e, d_{\alpha}d_{\beta}=d_{\beta}d_{\alpha}=d_{\alpha\beta} >, \label{eq:GD(G)ZnZnZnZ3} 
\end{align}
where $\tilde{\alpha} \equiv \alpha'\alpha^{\prime\prime -1}$, $\tilde{\alpha}' \equiv \alpha\alpha^{\prime -1}$, $d_X \equiv D(G)X$.
%\begin{align}
%\begin{array}{l}
%D(\Sigma(3n^3)) = < \tilde{\alpha} \equiv \alpha'\alpha^{\prime\prime -1}, \tilde{\alpha}' \equiv \alpha\alpha^{\prime -1} | \tilde{\alpha}^n=\tilde{\alpha}^{\prime n}=e, \tilde{\alpha}\tilde{\alpha}'=\tilde{\alpha}'\tilde{\alpha} > = Z_n \times Z_n, \\
%\Sigma(3n^3)/D(\Sigma(3n^3)) \simeq Z_n \times Z_3.
%\end{array}
%\label{DGDZnZnZnZ3}
%\end{align}
In the class (iii), we can obtain
\begin{align}
&G_0 = \left\{
\begin{array}{ll}
< \tilde{\alpha}, \tilde{\alpha}', \tilde{\beta} | \tilde{\beta}^3=(\tilde{\alpha}\tilde{\alpha}^{\prime -1})^{n/3}, \tilde{\beta}\tilde{\alpha}\tilde{\beta}^{-1}=\tilde{\alpha}^{-1}\tilde{\alpha}^{\prime -1}, \tilde{\beta}\tilde{\alpha}'\tilde{\beta}^{-1}=\tilde{\alpha} > \equiv
R_n & (n \in 3\mathbb{Z}) \\
< \tilde{\alpha}, \tilde{\alpha}' > =
Z_n \times Z_n & ({\rm otherwise})
\end{array}
\right., \\
&G/G_0 \simeq Z_N = \left\{
\begin{array}{ll}
Z_n = < g_{\alpha} | g_{\alpha}^n=g_e > & (n \in 3\mathbb{Z}) \\
Z_n \times Z_3 \simeq Z_{3n} = <g_{\tilde{\gamma}} | g_{\tilde{\gamma}}^{2n}=g_e > & ({\rm otherwise})
\end{array}
\right.,
\end{align}
%\begin{align}
%\begin{array}{l}
%&G_0 = \left\{
%\begin{array}{ll}
%< \tilde{\alpha}, \tilde{\alpha}', \tilde{\beta} | \tilde{\beta}^3=(\tilde{\alpha}\tilde{\alpha}^{\prime -1})^{n/3}, \tilde{\beta}\tilde{\alpha}\tilde{\beta}^{-1}=\tilde{\alpha}^{-1}\tilde{\alpha}^{\prime -1}, \tilde{\beta}\tilde{\alpha}'\tilde{\beta}^{-1}=\tilde{\alpha} > \equiv
%R_n & (n \in 3\mathbb{Z}) \\
%< \tilde{\alpha}, \tilde{\alpha}' > =
%Z_n \times Z_n & ({\rm otherwise})
%\end{array}
%\right., \notag \\
%&\Sigma(3n^3)/G_0 \simeq Z_N = \left\{
%\begin{array}{ll}
%Z_n & (n \in 3\mathbb{Z}) \\
%Z_n \times Z_3 \simeq Z_{3n} & ({\rm otherwise})
%\end{array}
%\right. 
%\end{array}
%\end{align}
where $\tilde{\beta} \equiv \alpha^{-n/3}\beta$, $\tilde{\gamma} \equiv \alpha^{-(n \mp 1)/3}\beta$, $g_X=G_0X$, and we omit relations between $\tilde{\alpha}$ and $\tilde{\alpha}'$ since they are same as Eq.~(\ref{eq:D(G)ZnZnZnZ3}).
We also introduce $\tilde{\alpha}'' \equiv \tilde{\gamma}^3$ and $\tilde{\beta}' \equiv \tilde{\gamma}^n$.
Here, $R_n$ is
%defined as the following.
%We define the generators as $\tilde{\alpha} \equiv \alpha\alpha^{\prime -1}$, $\tilde{\alpha}' \equiv \alpha'\alpha^{\prime\prime -1}$, and $\tilde{\beta} \equiv \alpha\beta$ for $\alpha \in Z_n$, $\alpha' \in Z'_n$, $\alpha'' \in Z''_n$, and $\beta \in Z_3$.
%By using these generators, $R_n$ is defined as
%\begin{align}
%R_n \equiv < \tilde{\alpha}, \ \tilde{\alpha}', \ \tilde{\beta} | \tilde{\alpha}^n=\tilde{\alpha}^{\prime n}=e, \ \tilde{\beta}^3=(\tilde{\alpha}\tilde{\alpha}^{\prime -1})^{n/3}, \ \tilde{\alpha}\tilde{\alpha}'=\tilde{\alpha}'\tilde{\alpha}, \ \tilde{\beta}\tilde{\alpha}\tilde{\beta}^{-1}=\tilde{\alpha}', \ \tilde{\beta}\tilde{\alpha}'\tilde{\beta}^{-1}=\tilde{\alpha}^{-1}\tilde{\alpha}^{\prime -1} >,
%\end{align}
%which is 
related to the following $\Delta(3n^2)$ as with the case that $Q_n$ is related to $D_n$.
Since $\alpha^n=\tilde{\gamma}^{3n}(=\tilde{\alpha}^{\prime\prime n}=\tilde{\beta}^{\prime 3})=e$, similarly,
$G=\Sigma(3n^3)$ can be written as
\begin{align}
\begin{array}{rl}
\Sigma(3n^3)=G
&\simeq G_0 \rtimes Z_N \\
&\simeq \left\{
\begin{array}{ll}
R_n \rtimes Z_n & (n \in 3\mathbb{Z}) \\
(Z_n \times Z_n) \rtimes Z_{3n} & ({\rm otherwise})
\end{array}
\right. \\
&\simeq \left\{
\begin{array}{ll}
R_n \rtimes Z_n & (n \in 3\mathbb{Z}) \\
((Z_n \times Z_n) \rtimes Z_{3}) \times Z_n \simeq \Delta(3n^2) \times Z_n & ({\rm otherwise})
\end{array}
\right..
\end{array}
\end{align}
Here, the structure of the semidirect product in the case of $n \in 3\mathbb{Z}$ comes from $\alpha \tilde{\beta} \alpha^{-1} = \tilde{\alpha}'\tilde{\beta}$, while the structures of semidirect and direct products in the case of $n \in 3\mathbb{Z} \pm 1$ come from $\tilde{\beta}'\tilde{\alpha}\tilde{\beta}^{\prime -1}=\tilde{\alpha}^{-1}\tilde{\alpha}^{\prime -1}$, $\tilde{\beta}'\tilde{\alpha}'\tilde{\beta}^{\prime -1}=\tilde{\alpha}$, $\tilde{\alpha}'' \delta \tilde{\alpha}^{\prime\prime -1} = \delta \ (\delta = \tilde{\alpha}, \tilde{\alpha}', \tilde{\beta}')$.

Next, let us see another example, $G=\Delta(3n^2) \simeq (Z_n \times Z'_n) \rtimes Z_3$.
When we take $\alpha \in Z_n$, $\alpha' \in Z'_n$, and $\beta \in Z_3$, 
they satisfy the following algebraic relations:
\begin{align}
\alpha^n = \alpha^{\prime n} = \beta^3 = e,
\end{align}
and also
\begin{align}
&\alpha' \alpha \alpha^{\prime -1} = \alpha \in Z_n, \quad (\alpha \alpha' \alpha^{-1} = \alpha' \in Z'_n) \\
&\beta \alpha \beta^{-1} = \alpha^{m_1} \alpha^{\prime m_2} \in Z_n \times Z'_n, \quad \beta \alpha^{-1} \beta^{-1} = \alpha^{m_3} \alpha^{\prime m_4} \in Z_n \times Z'_n. \label{eq:semidircond1ZnZnZ3}
\end{align}
Here, $m_i\ (i=1,2,3,4)$ can be determined  by the constraints, $\beta^3 \alpha \beta^{-3}=\alpha$, $\beta^3 \alpha' \beta^{-3}=\alpha'$,
%\begin{align}
%\alpha
%= \beta^3 \alpha \beta^{-3} 
%= \alpha^{m_1(m_1^2+m_2m_3)+m_2m_3(m_1+m_4)} \alpha'^{m_2(m_1^2+m_4^2+m_1m_4+m_2m_3)}, \\
%\alpha'
%= \beta^3 \alpha' \beta^{-3} 
%= \alpha^{m_3(m_1^2+m_4^2+m_1m_4+m_2m_3)} \alpha'^{m_4(m_4^2+m_2m_3)+m_2m_3(m_1+m_4)},
%\end{align}
and then,  we obtain $m_1=m_2=-1,\ m_3=1,\ m_4=0$.
Thus, Eq.~(\ref{eq:semidircond1ZnZnZ3}) can be rewritten as
\begin{align}
\beta \alpha \beta^{-1} = \alpha^{-1} \alpha^{\prime -1}, \quad \beta \alpha' \beta^{-1} = \alpha. \label{eq:semidircond2ZnZnZ3}
\end{align}
Then, we can obtain
\begin{align}
&D(G) = \left\{
\begin{array}{ll}
< \tilde{\alpha}, \tilde{\alpha}' | \tilde{\alpha}^n=\tilde{\alpha}^{\prime n/3}=e, \tilde{\alpha}\tilde{\alpha}'=\tilde{\alpha}'\tilde{\alpha} > = Z_n \times Z_{n/3} & (n \in 3\mathbb{Z}) \\
< \tilde{\alpha}, \tilde{\alpha}' | \tilde{\alpha}^n=\tilde{\alpha}^{\prime n}=e, \tilde{\alpha}\tilde{\alpha}'=\tilde{\alpha}'\tilde{\alpha} > = Z_n \times Z_n & ({\rm otherwise})
\end{array}
\right., \label{eq:D(G)ZnZnZ3} \\
&G/D(G) \simeq \left\{
\begin{array}{ll}
Z_3 \times Z_3 = < (d_{\alpha}, d_{\beta}) | d_{\alpha}^3=d_{\beta}^3=d_e, d_{\alpha}d_{\beta}=d_{\beta}d_{\alpha}=d_{\alpha\beta} > & (n \in 3\mathbb{Z}) \\
Z_3 = < d_{\beta} | d_{\beta}^3=d_e > & ({\rm otherwise})
\end{array}
\right., \label{eq:GD(G)ZnZnZ3}
\end{align}
where $\tilde{\alpha} \equiv \alpha \alpha^{\prime -1}$, $\tilde{\alpha}' \equiv \alpha^{-3}$, $d_X \equiv D(G)X$.
%\begin{align}
%\begin{array}{rl}
%D(\Delta(3n^2)) &= < \tilde{\alpha} \equiv \alpha \alpha^{\prime -1}, \tilde{\alpha}' \equiv \alpha^{-3} | \tilde{\alpha}^n=\tilde{\alpha}^{\prime n/3}=e, \tilde{\alpha}\tilde{\alpha}'=\tilde{\alpha}'\tilde{\alpha} >  \\
%&= \left\{
%\begin{array}{ll}
%Z_n \times Z_{n/3} & (n \in 3\mathbb{Z}) \\
%Z_n \times Z_n & ({\rm otherwise})
%\end{array}
%\right., \\
%\Delta(3n^2)/D(\Delta(3n^2)) &\simeq \left\{
%\begin{array}{ll}
%Z_3 \times Z_3 & (n \in 3\mathbb{Z}) \\
%Z_3 & ({\rm otherwise})
%\end{array}
%\right..
%\end{array}
%\label{DGDZnZnZ3}
%\end{align}

The above situation can also be understood in terms of $Z_n$ and $Z'_n$ charge constraints.
Let us assume that the chiral fermions are $Z_n$ and $Z'_n$ eigenstates, where the $[Z_n, Z'_n]$ charges of the $j$ th component field are $[q_j, q'_j]$.
Eq.~(\ref{eq:semidircond2ZnZnZ3}) shows that there exists a state with charge $[-(q_j+q'_j), q_j]$ and 
$\beta$ transforms the $j$ th component to the sate with charge $[-(q_j+q'_j), q_j]$.
When we consider a fundamental irreducible representation, it becomes a triplet with charge $ ^t([q_1, q'_1], [q_2, q'_2], [q_3, q'_3])$ $=$ $ ^t([q, q'], [-(q+q'), q], [q', -(q+q')])$, and then we have 
\begin{align}
{\rm det}\rho(\alpha) = {\rm det}\rho(\alpha') = 1, \label{eq:aa'}
\end{align}
which means Eqs.~(\ref{eq:D(G)ZnZnZ3}) and (\ref{eq:GD(G)ZnZnZ3}).
Note that for the triplet, Eq.~(\ref{eq:aa'}) and also ${\rm det}\rho(\beta)=1$ are satisfied, even if $n \in 3\mathbb{Z}$.

%Then, when we consider the same situation as Eq.~(\ref{eq:min}), 
In the class (iii), we can obtain
\begin{align}
&G_0 = \left\{
\begin{array}{ll}
< \tilde{\alpha}, \tilde{\alpha}', \tilde{\beta} | \tilde{\beta}^3=e, \tilde{\beta}\tilde{\alpha}\tilde{\beta}^{-1}=\tilde{\alpha}\tilde{\alpha}', \tilde{\beta}\tilde{\alpha}'\tilde{\beta}^{-1}=\tilde{\alpha}^{-3}\tilde{\alpha}^{\prime -2} > = (Z_n \times Z_{n/3}) \rtimes Z_3 & (n \in 3\mathbb{Z}) \\
< \tilde{\alpha}, \tilde{\alpha}' > = Z_n \times Z_n & ({\rm otherwise})
\end{array}
\right. \\
&G/G_0 \simeq Z_N = Z_3 = < g_{\beta} | g_{\beta}^3=g_e >, \quad \forall n,
\end{align}
%\begin{align}
%\begin{array}{rl}
%G_0
%&= \left\{
%\begin{array}{ll}
%< \tilde{\alpha}, \tilde{\alpha}', \tilde{\beta} \equiv \alpha\beta | \tilde{\beta}^3=e, \tilde{\beta}\tilde{\alpha}\tilde{\beta}^{-1}=\tilde{\alpha}\tilde{\alpha}', \tilde{\beta}\tilde{\alpha}'\tilde{\beta}^{-1}=\tilde{\alpha}^{-3}\tilde{\alpha}^{\prime -2} > & (n \in 3\mathbb{Z}) \\
%< \tilde{\alpha}, \tilde{\alpha}' > & ({\rm otherwise})
%\end{array}
%\right. \\
%&= \left\{
%\begin{array}{ll}
%(Z_n \times Z_{n/3}) \rtimes Z_3 & (n \in 3\mathbb{Z}) \\
%Z_n \times Z_n & ({\rm otherwise})
%\end{array}
%\right., \\
%\Delta(3n^2)/G_0 &\simeq Z_N = Z_3, \quad \forall n,
%\end{array}
%\end{align}
where $\tilde{\beta} \equiv \alpha^{-1}\beta$, $g_{\beta} \equiv G_0\beta$, and we omit relations between $\tilde{\alpha}$ and $\tilde{\alpha}'$ since they are same as Eq.~(\ref{eq:D(G)ZnZnZ3}).
Then, since $\beta^3=\tilde{\beta}^3=e$,
we can write
\begin{align}
\begin{array}{rl}
\Delta(3n^2)=G
&\simeq G_0 \rtimes Z_3 \\
&\simeq \left\{
\begin{array}{ll}
(D(\Delta(3n^2)) \rtimes Z_3) \rtimes Z_3 &  (n \in 3\mathbb{Z}) \\
D(\Delta(3n^2)) \rtimes Z_3 & ({\rm otherwise})
\end{array}
\right. \\
&\simeq \left\{
\begin{array}{ll}
((Z_n \times Z_{n/3}) \rtimes Z_3) \rtimes Z_3 &  (n \in 3\mathbb{Z}) \\
(Z_n \times Z_n) \rtimes Z_3 & ({\rm otherwise})
\end{array}
\right.,
\end{array}
\end{align}
where the last $Z_3$ semidirect product in the case of $n \in 3\mathbb{Z}$ comes from
$\beta\tilde{\alpha}\beta^{-1}=\tilde{\alpha}\tilde{\alpha}'$, $\beta\tilde{\alpha}'\beta^{-1}=\tilde{\alpha}^{-3}\tilde{\alpha}^{\prime -2}$, $\beta\tilde{\beta} \beta^{-1} = \tilde{\alpha}^{-1}\tilde{\alpha}^{\prime -1}\tilde{\beta}$ .
For example, the group $G=A_4 \simeq \Delta(12)$ is included in this case, and then
we can find that the $A_4$ flavor model corresponds to either the class (iii), $G_0=D(G)=Z_n \times Z'_n$, $G/G_0 \simeq Z_N=Z_3$, or the class (i), $G_0=G=A_4$.
Indeed, the $A_4$ symmetry has four irreducible representations, ${\bf 1}$, ${\bf 1'}$, ${\bf 1''}$, and ${\bf 3}$, 
which have ${\rm det}\rho_{\bf 1}(\beta)={\rm det}\rho_{\bf 3}(\beta)=1$, ${\rm det}\rho_{\bf 1'}(\beta)=e^{2\pi i/3}$ and  
${\rm det}\rho_{\bf 1''}(\beta)=e^{4\pi i/3}$~\cite{Ishimori:2010au,Ishimori:2012zz}.
The whole $A_4$ symmetry is anomaly free in flavor models including a proper number of ${\bf 1'}$ and ${\bf 1''}$. Otherwise, the $Z_3$ subsymmetry can be anomalous.
We note that, in the case of double covering group $T'$, there is no modification except double covering.

\end{step}

\begin{step}[$G \simeq K_G \rtimes Z_B$]
The above two types in the steps $1$ and $2$ are specific examples of $G \simeq K_G \rtimes Z_B$ type.
Then, let us consider a more generic case, $G \simeq K_G \rtimes Z_B\ (G/K_G \simeq Z_B)$.
In this case, it is found that $D(G) \subseteq K_G$.
%Even if we do not know the specific structure of $D(G)$,
Then, we can find that
\begin{align}
G/D(G) \simeq (K_G/D(G)) \times Z_B \label{eq:GD(G)KGZB}
\end{align}
from the following proof.
\begin{pf}
We can prove Eq.~(\ref{eq:GD(G)KGZB})
\begin{enumerate}
\renewcommand{\labelenumi}{(\Roman{enumi})}
\item From the isomorphism theorem $3$ in Appendix~\ref{app:isomorphism},
it is satisfied that 
\begin{align}
Z_B \simeq G/K_G \simeq (G/D(G))/(K_G/D(G)).
\end{align}
\item Since $G \simeq K_G \rtimes Z_B$, $Z_B$ is a subgroup of $G$ and 
the relation $K_G \cap Z_B = \{e\}$ is satisfied.
Then, the relation $(K_G/D(G)) \cap Z_B = \{e\}$ is also satisfied.
Thus, we can write $G/D(G) \simeq (K_G/D(G)) \rtimes Z_B$.
\item In particular, since $G/D(G)$ is Abelian, $G/D(G) \simeq (K_G/D(G)) \times Z_B$.
\end{enumerate}
\end{pf}
Then, in the class (iii), $G/G_0 \simeq Z_N$ can be obtained by calculating the least common multiple of all orders of cyclic groups. 
% even if  we do not know the specific structure of $G_0$.

As an example which is not mentioned in the steps $1$ and $2$, let us consider $G=S_n \simeq A_n \rtimes Z_2\ (n \geq 5)$.
Since $A_n\ (n \geq 5)$ is a perfect group, $A_n$ should be included in $D(G)$,  $A_n \subseteq D(G)$, 
and then we find $D(G)=A_n$.
Thus, we can find that the $S_n$ flavor model corresponds to either the class (iii), $G_0=D(G)=A_n$, $G/G_0 \simeq Z_N=Z_2$, or the class (i), $G_0=G=S_n$.
%In the class (i), we obtain $G_0=S_n$ and the the whole $S_n$ is anomaly free in particle theories.
%
\end{step}

\begin{step}[General $G \simeq K_G \rtimes G^{(1)}$]
Finally, let us consider general case $G \simeq K_G \rtimes G^{(1)}\ (G/K_G \simeq G^{(1)})$.
When $G^{(1)}$ is Abelian, it comes down to the step $3$ because of the fundamental structure theorem of finite Abelian group in Appendix~\ref{app:Abelian} and Eqs.~(\ref{eq:decomposiG1}) and (\ref{eq:decomposiK}).
Generally, as discussed in the step $3$, we can obtain
\begin{align}
G/D(G) \simeq (K_G/D(G)) \times G^{(1)}({\rm Abelian}).
\end{align}

Now, let us consider the case that $G^{(1)}$ is non-Abelian.
Since $G/K_G \simeq G^{(1)}$ and $G^{(1)} \triangleright D(G^{(1)}) \neq \{e\}$, by use of the correspondence theorem in Appendix~\ref{app:isomorphism}, the relation $K_G \subset D(G) \subseteq G_0$ is satisfied.
Here, similar to $G_0$, we define $G_0^{(1)} \equiv \{ g_0^{(1)} \in G^{(1)} | {\rm det}\rho(g_0^{(1)})=1 \}$, and then 
we see $D(G^{(1)}) \subseteq G_0^{(1)}$.
By use of the isomorphism theorem $3$ in Appendix~\ref{app:isomorphism}, we can find that
\begin{align}
\begin{array}{rl}
Z_N
&\simeq G/G_0 \\
&\simeq (G/D(G))/(G_0/D(G)) \\
&\simeq [(G/K_G)/(D(G)/K_G)]/[(G_0/K_G)/(D(G)/K_G)] \\
&\simeq (G^{(1)}/D(G^{(1)}))/(G_0^{(1)}/D(G^{(1)})) \\
&\simeq G^{(1)}/G_0^{(1)}.
\end{array}
\end{align}
Therefore, in this case, the structures of $D(G)$, $G/D(G)$, $G_0$, and $G/G_0$ depend on $D(G^{(1)})$, $G^{(1)}/D(G^{(1)})$, $G_0^{(1)}$, and $G^{(1)}/G_0^{(1)}$, respectively.

For example, let us consider $G=\Delta(6n^2)$,
\begin{align}
\begin{array}{rl}
\Delta(6n^2)
&\simeq (Z_n \times Z'_n) \rtimes S_3
\simeq (Z_n \times Z'_n) \rtimes (Z_3 \rtimes Z_2) \\
&\simeq ((Z_n \times Z'_n) \rtimes Z_3) \rtimes Z_2
\simeq \Delta(3n^2) \rtimes Z_2.
\end{array}
\end{align}
Then, we can find that $Z_n \times Z'_n \subset G_0$ and $Z_N \simeq S_3/G_0^{(1)}$.
In addition, it is satisfied that $S_3 \simeq A_3 \rtimes Z_2 \simeq Z_3 \rtimes Z_2$ and $G_0^{(1)} \supseteq A_3 \simeq Z_3$.
Thus, we can find that $\Delta(6n^2)$ flavor model corresponds to either the class (iii), $G_0=D(G)=\Delta(3n^2)$, $G/G_0 \simeq Z_N=Z_2$, or the class (i), $G_0=G=\Delta(6n^2)$.
For example, $S_4 \simeq \Delta(24) \simeq \Delta(12) \rtimes Z_2 \simeq A_4 \rtimes Z_2$ is included in this case.
Indeed, the $S_4$ symmetry has five irreducible representations, ${\bf 1}$, ${\bf 1'}$, ${\bf 2}$, ${\bf 3}$, and ${\bf 3'}$, 
which have ${\rm det}\rho_{\bf 1}(\beta)={\rm det}\rho_{\bf 3'}(\beta)=1$, ${\rm det}\rho_{\bf 1'}(\beta)={\rm det}\rho_{\bf 2}(\beta)={\rm det}\rho_{\bf 3}(\beta)=-1$~\cite{Ishimori:2010au,Ishimori:2012zz}.
The whole $S_4$ symmetry is anomaly free in flavor models including even numbers of ${\bf 1'}$, ${\bf 2}$, and ${\bf 3}$.
Otherwise, the $Z_2$ subsymmetry can be anomalous.
\end{step}

So far, we have seen the detail structure of the anomaly-free subgroup $G_0$ and the anomalous part $G/G_0 \simeq Z_N$ for various typical groups $G$ from the structure of the derived subgroup $D(G)$ and the residue class group $G/D(G)$.
Here, we list $D(G)$ of those typical groups $G$ and $G/D(G)$.
We note again that the derived subgroup $D(G)$ is also automatically the subgroup of anomaly-free group $G_0$, which holds in any arbitrary representations.
From the following Table~\ref{tab:D(G)GD(G)}, when we consider $G=S_n \simeq A_n \rtimes Z_2$ and $G=\Delta(6n^2) \simeq \Delta(3n^2) \rtimes Z_2$, in particular, we can find that $G_0 \supseteq A_n$ and $G_0 \supseteq \Delta(3n^2)$ are at least satisfied, respectively.
\begin{table}[H]
\centering
\begin{tabular}{|c|c|c|} \hline
$G$ & $D(G) (\subseteq G_0)$ & $G/D(G)$ \\ \hline \hline
$Z_p$ & $\{e\}$ & $Z_p$ \\
$(A_3 \simeq Z_3)$ & $(\{e\}$) & $(Z_3 \simeq A_3)$ \\
\hline
$D_n \simeq Z_n \rtimes Z_2$ &
$\left\{
\begin{array}{cc}
Z_{n/2} & (n \in 2\mathbb{Z}) \\
Z_n & (n \in 2\mathbb{Z}+1)
\end{array}
\right.$ &
$\left\{
\begin{array}{cc}
Z_2 \times Z_2 & (n \in 2\mathbb{Z}) \\
Z_2 & (n \in 2\mathbb{Z}+1)
\end{array}
\right.$ \\
$(S_3 \simeq D_3 \simeq A_3 \rtimes Z_2)$ & $(Z_3 \simeq A_3)$ & $(Z_2)$ \\
\hline
$T_{p^k} \simeq Z_{p^k} \rtimes Z_3\ (p \neq 3)$ & $Z_{p^k}$ & $Z_3$ \\
\hline
$\Sigma(2n^2) \simeq (Z_n \times Z_n) \rtimes Z_2$ & $Z_n$ & $Z_n \times Z_2$ \\
\hline
$\Sigma(3n^3) \simeq (Z_n \times Z_n \times Z_n) \rtimes Z_3$ & $Z_n \times Z_n$ & $Z_n \times Z_3$ \\
\hline
$\Delta(3n^2) \simeq (Z_n \times Z_n) \rtimes Z_3$ & 
$\left\{
\begin{array}{cc}
Z_n \times Z_{n/3} & (n \in 3\mathbb{Z}) \\
Z_n \times Z_n & ({\rm otherwise})
\end{array}
\right.$ &
$\left\{
\begin{array}{cc}
Z_3 \times Z_3 & (n \in 3\mathbb{Z}) \\
Z_3 & ({\rm otherwise})
\end{array}
\right.$ \\
$(A_4 \simeq \Delta(12))$ & $(Z_2 \times Z_2)$ & $(Z_3)$ \\
\hline
$\begin{array}{rl}
\Delta(6n^2)
&\simeq (Z_n \times Z_n) \rtimes S_3 \\
&\simeq (Z_n \times Z_n) \rtimes (Z_3 \rtimes Z_2) \\
&\simeq ((Z_n \times Z_n) \rtimes Z_3) \rtimes Z_2 \\
&\simeq \Delta(3n^2) \rtimes Z_2
\end{array}$ & $\Delta(3n^2)$ & $Z_2$ \\
$\begin{array}{rl}
(S_4
&\simeq \Delta(24) \\
&\simeq \Delta(12) \rtimes Z_2 \\
&\simeq A_4 \rtimes Z_2)
\end{array}$
%$(S_4 \simeq \Delta(24) \simeq \Delta(12) \rtimes Z_2 \simeq A_4 \rtimes Z_2)$
& $(\Delta(12) \simeq A_4)$ & $(Z_2)$ \\
\hline
$PSL(2,Z_p)\ (p \neq 2,3)$ & $PSL(2,Z_p)\ (p \neq 2,3)$ & - \\
\hline
$A_n\ (n \geq 5)$ & $A_n\ (n \geq 5)$ & - \\
\hline
$S_n \simeq A_n \rtimes Z_2\ (n \geq 5)$ & $A_n\ (n \geq 5)$ & $Z_2$ \\
\hline
\end{tabular}
\caption{The derived subgroup $D(G)$ of typical groups $G$ and the residue class group $G/D(G)$.}
\label{tab:D(G)GD(G)}
\end{table}

% ------------------------------------------------------ %
% ------------------------------------------------------ %

Finally, let us summarize the important points in this section again.
\begin{itemize}
\item
The derived subgroup of $G$, $D(G)$, in Eq.~(\ref{eq:D(G)}) is always included in $G_0$,~i.e., $D(G) \subseteq G_0\ (D(G) \triangleleft G_0)$.
It does not depend on representations of $G$.
While $G/G_0 \simeq Z_N$, $G/D(G)$ becomes Abelian in Eq.~(\ref{eq:generalGD(G)}) and each cyclic groups can be anomalous.
The number $N$ can be found by the least common multiple of orders of the anomalous cyclic subgroups.
Thus, $D(G)$ gives an important clue to obtain information about the anomaly-free subgroup $G_0$ and the anomalous part $G/G_0 \simeq Z_N$.
\item
The detail structure of $D(G)$ depends on the structure of $G$.
However, if $G$ can be written as $G \simeq K_G \rtimes G^{(1)}$, we can obtain some information as discussed in the above step $4$; in the case that $G^{(1)}$ is Abelian, 
the relations $D(G) \subseteq K_G$ and $G/D(G) \simeq (K_G/D(G)) \times G^{(1)}$ are satisfied, while in the case that $G^{(1)}$ is non-Abelian, the relations $K_G \subset D(G) \subseteq G_0$ and $G/D(G) \simeq G^{(1)}/D(G^{(1)})$ are satisfied.
In particular, when we consider $G=S_n \simeq A_n \rtimes Z_2$ and $G=\Delta(6n^2) \simeq \Delta(3n^2) \rtimes Z_2$, 
we find $G_0 \supseteq A_n$ and $G_0 \supseteq \Delta(3n^2)$ at least.
\end{itemize}

% ------------------------------------------------------ %
% ------------------------------------------------------ %
% ------------------------------------------------------ %
% ------------------------------------------------------ %

\section{Comment on generic theories with $M>1$}
\label{sec:comment}

We have concentrated on the theory 
with $\sum_{\bf R}2T_2({\bf R})=1$ and the anomaly-free condition 
 ${\rm det}\rho (g) =1$.
In this theory, the anomaly-free elements correspond to a normal subgroup $G_0$ of $G$.
Then, the residue class group $G/G_0 \simeq Z_N$ can be anomalous.
We can extend our analysis to  the theory with $\sum_{\bf R}2T_2({\bf R})=M > 1$
 and 
anomaly-free condition $({\rm det}\rho (g))^M =1$.

When the determinant of the representation of any element $\forall g \in G$ satisfies $({\rm det}\rho(g))^N=1$, that of the anomaly-free element $g_n$ satisfies $({\rm det}\rho(g_n))^n=1$ with $n={\rm gcd}(N,M)$, which means the determinant can be written as ${\rm det}\rho(g_n)=e^{2\pi iQ''(g_n)/n}$.
Then, we define the subset of $G$,
\begin{align}
G_n \equiv \{ g_n \in G | {\rm det}\rho(g_n) = e^{2\pi i Q'(g_n)/N} = e^{2\pi i Q''(g_n)/n} \}, 
\end{align}
where $Q'(g_n)(=Q(g_n)N/N(g_n))=Q''(g_n)N/n$.
Similar to $G_0$, we find that 
$G_n$ is also a normal subgroup of $G$, $G_n \triangleleft G$.
In addition, $G_n$ includes of $G_0$, $G_0 \subset G_n$.
That also means $G_0 \triangleleft G_n$.
Then, we can similarly derive $G/G_n \simeq Z_{N/n}$ and $G_n/G_0 \simeq Z_n$. Indeed, by use of the isomorphism theorem $3$ in Appendix~\ref{app:isomorphism}, we find that
\begin{align}
G/G_n \simeq (G/G_0)/(G_n/G_0) \simeq Z_N/Z_n \simeq Z_{N/n}.
\end{align}
That is, in this particle theory, the subgroup $G_n$ is always anomaly free, but 
$ Z_{N/n}$ can be anomalous.
Thus, the anomalous symmetry is the single cyclic group again.
Furthermore, if there exists $\exists g \in G$ which satisfies $N(g)=N/n$ and ${\rm gcd}(Q(g),N(g))=1$, $G$ can be expressed as
\begin{align}
G \simeq G_n \rtimes Z_{N/n}.
\end{align}
In this case, the anomaly-free and anomalous parts of $G$ can be separated.

There is an interesting example.
When $M$ is a multiple of $N$, the whole symmetry $G$ is anomaly free.
As the example, let us assume that our particle theory has $E_6$ gauge symmetry\footnote{$E_6$ gauge symmetry is automatically anomaly free.} and all chiral fermions in the theory transform as ${\bf 27}^i$ representation under $E_6$ transformation, where $i$ denotes the flavor index.
(For example, the $i$ th generational standard model quarks and leptons are embedded in ${\bf 27}^i$ representation.)
Furthermore, we also assume non-Abelian discrete flavor symmetry $G$ among those chiral fermions at least at classical level.
In this case, we find that $M=2T_2({\bf 27})=6$.
Then, from Eq.~(\ref{eq:GG0iii}), when $G$ corresponds to either of the groups listed in the Table~\ref{tab:D(G)GD(G)} except $\Sigma(2n^2)$ and $\Sigma(3n^2)$, at least, we can find that whole $G$ flavor symmetry can be always anomaly free whatever the fermions have any representations of $G$.

% ------------------------------------------------------ %
% ------------------------------------------------------ %
% ------------------------------------------------------ %
% ------------------------------------------------------ %

\section{Conclusion}
\label{sec:conclusion}

We have studied the anomaly-free subgroup $G_0$ of a discrete group $G$.
If the determinant of a chiral transformation is trivial, ${\rm det}\rho(g_0)=1\ (g_0 \in G)$, in particular, the transformation is anomaly free.
We have found that the anomaly-free transformations generate a normal subgroup of the group $G$, $G_0 \triangleleft G$, and the residue class group $G/G_0$, which becomes the anomalous part of $G$, is isomorphic to a single cyclic group $Z_N$, where the number $N$ can be read from ${\rm det}\rho(g)=e^{2\pi iQ'(g)/N}\ (\forall g \in G)$.
In particular, if this $Z_N$ actually becomes the subgroup of $G$, $G$ can be written as $G \simeq G_0 \rtimes Z_N$, which means that the anomaly-free and anomalous parts can be separated.
Furthermore, this structure constrains the structure of $G_0$; $G_0$ has to contain the derived subgroup $D(G)$ in Eq.~(\ref{eq:D(G)}) due to $G/G_0 \simeq Z_N$, and $D(G)$ is also a normal subgroup of $G_0$.
In addition, the number $N$ is also constrained as a divisor of the least common multiple of orders of each cyclic group of $G/D(G)$.
In this sense, $D(G)$ and $G/D(G)$ give an important clue to obtain information about the anomaly-free subgroup $G_0$ and the anomalous part $G/G_0 \simeq Z_N$.

Then, we have studied their detail structure in various discrete groups $G$.
In particular, if $G$ can be written as $G \simeq K_G \rtimes G^{(1)}$, we can obtain some information; in the case that $G^{(1)}$ is Abelian, we find $D(G) \subseteq K_G$ and $G/D(G) \simeq (K_G/D(G)) \times G^{(1)}$, while in the case that $G^{(1)}$ is non-Abelian, we find $K_G \subset D(G) \subseteq G_0$ and $G/D(G) \simeq G^{(1)}/D(G^{(1)})$.
Interestingly, when we consider $G=S_n \simeq A_n \rtimes Z_2$ and $G=\Delta(6n^2) \simeq \Delta(3n^2) \rtimes Z_2$, 
we obtain $G_0 \supseteq A_n$ and $G_0 \supseteq \Delta(3n^2)$ at least.
This result holds in any arbitrary representations.
We would like to study some applications elsewhere.

These analyses can be extended to particle theories with $\sum_{\bf R}2T_2({\bf R})=M > 1$.
Then, the group $G$ has the anomaly-free normal subgroup $G_n$ and then the residue class group $G/G_n \simeq Z_{N/n}$ becomes the anomalous part, where $n={\rm gcd}(N,M)$.
Similarly, if this $Z_{N/n}$ actually becomes the subgroup of $G$, $G$ can be written as $G \simeq G_n \rtimes Z_{N/n}$.
Interestingly, if we consider that the particle theory has $E_6$ gauge symmetry and a discrete flavor symmetry $G$, and all chiral fermions in the theory are represented as {\bf 27} under $E_6$, at least most of groups $G$ in Table~\ref{tab:D(G)GD(G)} can be always anomaly free whatever the fermions have any representations of $G$.
%Then, the symmetry $G$ is decomposed into anomaly-free one $G_n$ and the anomalous one $Z_{N/n}$, i.e. 
%$G \simeq G_n \rtimes Z_{N/n}$, where $n={\rm gcd}(N,M)$.
%We have concentrated on the theory 
%with $\sum_{\bf R}2T_2({\bf R})=1$ and the anomaly-free condition 
% ${\rm det}\rho (g) =1$.
%In this theory, the anomaly-free elements correspond to a normal subgroup $G_0$ of $G$.
%Then, the residue class group $G/G_0 \simeq Z_N$ can be anomalous.
%We can extend our analysis to  the theory with $\sum_{\bf R}2T_2({\bf R})=M > 1$ and 
%anomaly-free condition $({\rm det}\rho (g))^M =1$.
%In this theory, the anomaly-free elements construct a subgroup $G_0^{(M)}$, where 
%$({\rm det}\rho (g))^M =1$ for $\forall g\in G$.
%Then, the group $G/G_0^{(M)}=Z_K$ can be anomalous, where 
%$K=N/{\rm gcd}(N,M)$.
%For example, when $M$ is a multiple of $N$, the whole symmetry $G$ is anomaly free.

We comment the Green-Schwarz mechanism for anomaly cancellation in four-dimensional low-energy effective 
field theory derived from superstring theory.
(See for review Refs.~\cite{Blumenhagen:2006ci,Ibanez}.)
Some of $U(1)$ gauge symmetries are anomalous in such effective field theory,  but 
those anomalies can be canceled by the Green-Schwarz mechanism, where axions shift 
under the anomalous $U(1)$ symmetries.
If the cyclic symmetry $G/G_0 \simeq Z_N$ can be embedded into those anomalous $U(1)$ symmetries, 
anomalies of the $Z_N$ symmetry can also be canceled by the same Green-Schwarz mechanism\footnote{
See for examples Refs.~\cite{Chen:2015aba,Hamada:2014hpa,Kariyazono:2019ehj}.}.
From this point, it would be important that only the single $Z_N$ symmetry is anomalous, but not 
generic Abelian group.

%-------- acknowledgement -------%
\vspace{1.5 cm}
\noindent
{\large\bf Acknowledgement}\\

Authors would like to thank S.~Kikuchi, K.~Nasu, and S.~Takada for useful discussions.
H. U. was supported by Grant-in-Aid for JSPS Research Fellows No. JP20J20388.

%-------- Appendix -------%

\appendix

\section{Isomorphism Theorems}
\label{app:isomorphism}

We introduce the fundamental homomorphism and then the isomorphism theorems as well as the correspondence theorem.
\begin{thmhom}
Let $K$ be a normal subgroup of $G$, $K \triangleleft G$, and then there is a natural homomorphism $\pi: G \rightarrow G/K$.
Let $f: G \rightarrow G'$ be a group homomorphism. In this case, ${\rm Ker}(f) \triangleleft G$ and we can consider $G'={\rm Im}(f)$ without loss of generality.
If $K$ is a subset of ${\rm Ker}(f)$, $K \subseteq {\rm Ker}(f)$, there exists a unique homomorphism $F: G/K \rightarrow G'$ such that $F \circ \pi = f$.
\end{thmhom}
\begin{thmiso}
In particular, if $K={\rm Ker}(f)$, $F: G/K \rightarrow G'(={\rm Im}(f))$ becomes a isomorphism;
\begin{align}
G/{\rm Ker}(f) \simeq {\rm Im}(f).
\end{align}
This is always satisfied when we consider $f: G \rightarrow G'$.
\end{thmiso}
\begin{thmcor}
Then, it can be applied even if $K \subset {\rm Ker}(f)$.
In this case, $G/K \triangleright {\rm Ker}(F) \neq \{e\}$ and then
\begin{align}
(G/K)/{\rm Ker}(F) \simeq G' \simeq G/{\rm Ker}(f)
\end{align}
is satisfied.
% and then $G/{\rm Ker}(f) \simeq (G/N)/{\rm Ker}(F)$.
Here, $K \subset {\rm Ker}(f) = {\rm Ker}(F \circ \pi) = \pi^{-1}({\rm Ker}(F)) \triangleleft G$.
% and then $N \subset \pi^{-1}({\rm Ker}(F)) \triangleleft G$.
Furthermore, there exists a group homomorphism $\phi: G \rightarrow G''$, which satisfies ${\rm Ker}(\phi)=K$ and then $G'' \simeq G/K$.
Accordingly, there exists $\tilde{K}'' \triangleleft G''$ such that $\tilde{K}'' \simeq {\rm Ker}(F)$.
Thus, for $\phi: G \rightarrow G''$,
\begin{align}
G/\tilde{K} \simeq G''/\tilde{K}''
\end{align}
is satisfied in general, where $\tilde{K}'' \triangleleft G''$ and $\tilde{K} \equiv \phi^{-1}(\tilde{K}'') (\supset {\rm Ker}(\phi)) \triangleleft G$.
This is often called the corresponding theorem.
%Thus, by taking $\tilde{N} \equiv \phi^{-1}(\tilde{N}'') \supset N = {\rm Ker}(\phi)$, $\tilde{N} \triangleleft G$ and $G/\tilde{N} \simeq G''/\tilde{N}''$ is satisfied in general.
\end{thmcor}
\begin{thmiso}
Let $K$ be a normal subgroup of $G$, $K \triangleleft G$. Let $H$ be a subgroup of $G$, $H \subset G$.
In this case, $K$ is also a normal subgroup of $KH$. In particular, $H \cap K$ is also a normal subgroup of $H$.
When we consider $\phi: KH \rightarrow H$, $\phi(K)=H \cap K$ and then
\begin{align}
KH/K \simeq H/(H \cap K)
\end{align}
is satisfied. 
\end{thmiso}
\begin{thmiso}
Let both $K_1$ and $K_2$ be normal subgroups of $G$ which satisfy $K_1 \subset K_2$.
In this case, $K_1 \triangleleft K_2$ and $K_2/K_1 \triangleleft G/K_1$.
By considering the corresponding theorem,
\begin{align}
G/K_2 \simeq (G/K_1)/(K_2/K_1)
\end{align}
is satisfied.
\end{thmiso}

\section{Semidirect Product}
\label{app:semidirect}

We comment on semidirect product.
If a normal subgroup of $G$, $K_G$, and a subgroup of $G$, $G^{(1)}$ satisfy the following conditions,
\begin{align}
G=K_G G^{(1)}, \quad K_G \cap G^{(1)} = \{e\}, \label{eq:condsemidirect}
\end{align}
$G$ can be expressed as
\begin{align}
G \simeq K_G \rtimes G^{(1)}.\label{eq:semidirect}
\end{align}
In particular, if any elements of $G^{(1)}$ commute all elements of $K_G$, $G$ can be expressed as $G \simeq K_G \times G^{(1)}$.
Applying the second isomorphism theorem in Appendix~\ref{app:isomorphism} to this, we can find $G/K_G \simeq G^{(1)}$.
We note that $G$ cannot be always expressed as $G \simeq K_G \rtimes G^{(1)}$ just because $G/K_G \simeq G^{(1)}$.
If this $G^{(1)}$ is actually a subgroup of $G$, Eq.~(\ref{eq:condsemidirect}) is satisfied and then $G$ can be expressed as Eq.~(\ref{eq:semidirect}).
In terms of components of Eq.~(\ref{eq:semidirect}), since $K_G$ is the normal subgroup of $G$,
\begin{align}
g_1 k g_1^{-1} = k^{(g_1)} \in K_G,
\end{align}
should be satisfied, where $k, k^{(g_1)} \in K_G$ and $g_1 \in G^{(1)}$.

Now, let us see the case that $G^{(1)}$ is further expressed as $G^{(1)} \simeq K_{G^{(1)}} \rtimes G^{(2)}$.
In this case, $G$ can be expressed as
\begin{align}
&G \simeq K_G \rtimes G^{(1)} \simeq K_G \rtimes (K_{G^{(1)}} \rtimes G^{(2)}) \label{eq:decomposiG1} \\
\Rightarrow \quad
&G \simeq (K_G \rtimes K_{G^{(1)}}) \rtimes G^{(2)} \simeq K'_{G} \rtimes G^{(2)}. \label{eq:decomposiK}
\end{align}
This can be found as the following by considering relations of their elements,
$k, k^{(k_1)}, k^{(g_2)}, k^{(g_2 \rightarrow k_1)} \in K_G$, $k_1, k_1^{(g_2)}, k_1^{(g_2 \rightarrow k)} \in K_{G^{(1)}}$, and $g_2 \in G^{(2)}$.
In the case of Eq.~(\ref{eq:decomposiG1}),
\begin{align}
k_1 k k_1^{-1} = k^{(k_1)}, \quad g_2 k g_2^{-1} = k^{(g_2)}, \quad g_2 k_1 g_2^{-1} = k_1^{(g_2)} \label{eq:relG1}
\end{align}
are satisfied, while in the case of Eq.~(\ref{eq:decomposiK}),
\begin{align}
k_1 k k_1^{-1} = k^{(k_1)}, \quad g_2 k g_2^{-1} = k^{(g_2)} k_1^{(g_2 \rightarrow k)}, \quad g_2 k_1 g_2^{-1} = k^{(g_2 \rightarrow k_1)} k_1^{(g_2)} \label{eq:relK}
\end{align}
are satisfied.
Thus, Eq.~(\ref{eq:relG1}) is sufficient condition for Eq.~(\ref{eq:relK}), but Eq.~(\ref{eq:relK}) does not always satisfy Eq.~(\ref{eq:relG1}).

\section{Finite Abelian Groups}
\label{app:Abelian}

We introduce some theorems of finite Abelian groups.
\begin{thmAbel}
Every finite Abelian group $G$ whose order $|G|=p_1^{A_1} \cdots p_r^{A_r}=\prod_{i=1}^{r} p_i^{A_i}$ can be expressed as
\begin{align}
G \simeq
(Z_{p_1^{a_{1,1}}}
%\times Z_{p_1^{a_{1,2}}}
\times \cdots \times Z_{p_1^{a_{1,n_1}}}) \times
%Z_{p_2^{a_{2,1}}} \times Z_{p_2^{a_{2,2}}} \times \cdots \times Z_{p_2^{a_{2,n_2}}}
\cdots \times
(Z_{p_r^{a_{r,1}}}
%\times Z_{p_r^{a_{r,2}}}
\times \cdots \times Z_{p_r^{a_{r,n_r}}}),
\end{align}
where each $p_i$ is a distinct prime number and $a_{i,j}$ satisfy
\begin{align}
A_i = \sum_{j=1}^{n_i} a_{i,j}, \quad a_{i,j} \geq a_{i,j+1}.
\end{align}
Note that $a_{i,j}$ is uniquely determined by $G$.
\end{thmAbel}
The above theorem uses the following theorem,
\begin{thmcirc}
If $m$ and $n$ are coprime to each other,
\begin{align}
Z_{mn} \simeq Z_{m} \times Z_{n} \label{eq:mn}
\end{align}
is satisfied.
\end{thmcirc}
Note that by using this theorem, the following relation,
\begin{align}
Z_m \times Z_n \simeq Z_{{\rm gcd}(m,n)} \times Z_{{\rm lcm}(m,n)} \label{eq:gcdlcm}
\end{align}
is generally satisfied.

%%%%%%%%%%%%%%%%%%%%%%%%%%%%%%%%%%%%%%%%%

\end{document}